\documentclass[twocolumn]{aastex631}
\usepackage{newtxtext,newtxmath}
\usepackage[T1]{fontenc}

\usepackage{graphicx,graphics}	
\usepackage{color,epsfig}
\usepackage{amsmath}	
\usepackage{amssymb}	
\usepackage{psfrag}
\usepackage[T1]{fontenc}
\usepackage{ebgaramond}
\usepackage{newtxmath}
\usepackage{booktabs} 
\usepackage{longtable}
\usepackage{supertabular}

\def\n{\hat{\bm{n}}}

\newcommand{\kk}{\boldsymbol{k}}
\newcommand{\kp}{\kk_\perp}
\newcommand{\kll}{k_\parallel} 

\newcommand{\U}{\boldsymbol{U}}

\newcommand{\bfl}{\boldsymbol{\ell}}

\newcommand{\Hi}{H\,\textsc{i}}
\newcommand{\HI}{H\,\textsc{i}~}

\def\a{{\bm{a}}}

\def\v{\bm{v}}

\def\V{\mathcal{V}}
\def\N{\mathcal{N}}
\def\S{{\mathcal S}}
\def\u{{\bf U}} 

\def\pp{\hat{\boldsymbol{\rm p}}}

\defcitealias{Gill_2024_2d3vc}{Paper~I}
\defcitealias{Gill_2025_mabs}{Paper~II}

\begin{document}

\title{ The  EoR 21-cm  Bispectrum at $z=8.2$ from MWA data I: Foregrounds and preliminary upper limits}

\correspondingauthor{Sukhdeep Singh Gill}
\email{sukhdeepsingh5ab@gmail.com}

\author[0000-0003-1629-3357]{Sukhdeep Singh Gill}
\affiliation{Department of Physics, Indian Institute of Technology Kharagpur, Kharagpur 721 302, India}

\author[0000-0002-2350-3669]{Somnath Bharadwaj}
\affiliation{Department of Physics, Indian Institute of Technology Kharagpur, Kharagpur 721 302, India}

\author[0000-0003-1206-8689]{Khandakar Md Asif Elahi}
\affiliation{Centre for Strings, Gravitation and Cosmology, Department of Physics, Indian Institute of Technology Madras, Chennai 600036, India}

\author[0000-0002-7737-1470]{Shiv K. Sethi}
\affiliation{Raman Research Institute, C. V. Raman Avenue, Sadashivanagar, Bengaluru 560080, India}

\author[0000-0002-6216-2430]{Akash Kumar Patwa}
\affiliation{Raman Research Institute, C. V. Raman Avenue, Sadashivanagar, Bengaluru 560080, India}

\begin{abstract}
We attempt to measure the $z = 8.2$ Epoch of Reionization (EoR)  21-cm bispectrum (BS) using Murchison Widefield Array (MWA)  $154.2~\mathrm{MHz}$ data.  We find that 
  $B(k_{1\perp}, k_{2\perp}, k_{3\perp}, k_{1\parallel}, k_{2\parallel})$
 the 3D cylindrical BS  exhibits a foreground wedge, similar to $P(k_{1\perp},k_{1\parallel})$  the 21-cm cylindrical power spectrum. However, the BS foreground wedge, which 
 depends on $(k_{1\perp},k_{1\parallel})$, $(k_{2\perp},k_{2\parallel})$ and $(k_{3\perp},k_{3\parallel})$ the three sides of a triangle, is more complicated.
  Considering various foreground avoidance scenarios, we identify the region where all three sides are outside the foreground wedge as the EoR window for the 21-cm BS. However, the EoR window is contaminated by a periodic pattern of spikes that arises from the periodic pattern of missing frequency channels in the data. We evaluate the binned 3D spherical BS for triangles of all possible sizes and shapes, and present results for $\Delta^3$ the mean cube brightness temperature fluctuations. The best $2\sigma$ upper limits we obtain for the EoR 21-cm signal are  
  $\Delta^3_{\rm UL} = (1.81\times 10^3)^3~\mathrm{mK}^3$ at $k_1 = 0.008~\mathrm{Mpc}^{-1}$ and $\Delta^3_{\rm UL} = (2.04\times 10^3)^3~\mathrm{mK}^3$ at $k_1 = 0.012~\mathrm{Mpc}^{-1}$ for equilateral and squeezed triangles, respectively. These are foreground-dominated, and are many orders of magnitude larger than the predicted EoR 21-cm signal  $(\sim 10^3 ~\mathrm{mK}^3)$. 
\end{abstract}



\section{Introduction}

The Epoch of Reionization (EoR) marks the phase transition of the Universe when the first luminous sources re-ionized the diffuse neutral Hydrogen (\Hi\relax) in the Inter-Galactic Medium (IGM). There are numerous indirect observations, such as the Cosmic Microwave Background (CMB) (\citealt{Planck_2016,Komatsu_2011,Larson_2011}), the absorption spectra of quasars (\citealt{Becker_2001,Fan_2003,White_2003,Barnett_2017}), and the luminosity function of Lyman-$\alpha$ emitters ( \citealt{Trenti_2010,Ouchi_2010,Jensen_2013,Choudhury_2015,Bouwens_2016,Ota_2017}), that constrain the EoR.  Nonetheless, our understanding of the physical processes occurring during the EoR remains quite limited, and the precise duration and mechanism are very little constrained by these observations \citep{Mitra_2015}. Observation of the redshifted  \Hi~21-cm radiation is a very promising probe of the EoR \citep{Loeb:2003ya,Bharadwaj_2004,Furlanetto_2006b,Pritchard_2012,Zaroubi_2013}, and  considerable efforts are underway to carry out such observations using radio telescopes such as the upgraded Giant Metrewave Radio Telescope (uGMRT)\footnote{\url{http://www.gmrt.ncra.tifr.res.in}} \citep{Swarup_1991,Gupta_2017}, Murchison Widefield Array
(MWA)\footnote{\url{https://www.mwatelescope.org/}} \citep{tingay13}, Low Frequency Array (LOFAR)\footnote{\url{https://www.astron.nl/telescopes/lofar/}} \citep{haarlem}, 
and Hydrogen
Epoch of Reionization Array (HERA)\footnote{\url{https://reionization.org/}} \citep{DeBoer_2017} , and the upcoming Square Kilometre Array (SKA)-LOW\footnote{\url{http://www.skatelescope.org}} \citep{Koopmans_2015}. 
The major obstacle for these observations lies in foregrounds that are 4-5 orders of magnitude larger than the expected 21-cm signal \citep{ali}. Currently, the best upper limits on the EoR 21-cm mean-squared brightness temperature fluctuations come from the HERA experiment, which are $[\Delta^2]\leq(30.76)^2$ mK$^2$ at $k = 0.192$ $h$ Mpc$^{ -1}$ at $z = 7.9$, and $[\Delta^2]\leq(95.74)^2$ mK$^2$ at $k = 0.256$  $h$ Mpc$^{ -1}$ $z = 10.4$ \citep{hera_2022}.

Much attention has been focused on the 21-cm power spectrum (PS),  or equivalently, on $[\Delta^2]$ the mean-squared brightness temperature fluctuations mentioned earlier.  Although these are adequate to completely quantify the statistics of a Gaussian random field, the 21-cm EoR signal is predicted to be highly non-Gaussian \citep{BP2005, mondal2015}, and it is necessary to consider higher-order statistics such as the bispectrum (BS). 
In an early work, \cite{BP2005} used a simple analytical model for the ionized bubble distribution to calculate the EoR 21-cm BS, which is predicted to be negative.
Since then, the EoR 21-cm BS has been extensively studied using simulated  \Hi~21-cm maps 
\citep{Yoshiura_2015, Shimabukuro_2016,Majumdar_2018, trott2019,Hutter_2019, Majumdar_2020, Saxena_2020, Watkinson_2021,Murmu:2023fgv, gill_eormulti, raste_2023}. 
The measurements of the EoR 21-cm BS can be used to tighten the constraints on the EoR models \citep{watkinson_2022,tiwari_2022},  and probe the IGM physics \citep{Watkinson_2021,Kamran_2021b, Kamran_2021,  kamran_2022}. Moreover, the BS will be crucial in validating any detection of the EoR 21-cm PS. In addition to the BS, the skewness and kurtosis have also been proposed to quantify the level of non-Gaussianity \citep{Harker_2009,Watkinson_2015,Shimabukuro_2015}, but being one-point statistics, they only capture limited information. The skew-spectrum has been proposed as an alternative to the BS to study the non-Gaussianity in the  EoR 21-cm signal \citep{Ma:2023eml}.
It is also possible to quantify non-Gaussianity by studying the geometry and topology of the EoR \Hi~ distribution \citep{Bag_2018,Bag_2019}.

To the best of our knowledge, there has been only one previous work that aims to measure the EoR 21-cm BS. \citet{trott2019} analyzed 21 hours of $167$–$197$ MHz  ($z = 6.2$ – $7.5$) MWA Phase II EoR project data to estimate the 21-cm BS for a few triangle configurations.
They found that for large-scale isosceles triangles, the thermal noise limit is achieved in 10 hours of observation, whereas the equilateral configurations are strongly foreground-dominated, with measured values several orders of magnitude above the noise predictions. The upper limit on the BS is measured to be $\sim 10^{12} ~\mathrm{mK}^3~\mathrm{Mpc}^6$ on large scales. The authors argue that by choosing triangle configurations that minimize the foreground bias, the BS detections could be achieved with only a few hundred hours of the MWA observations, potentially requiring less time than the equivalent PS detection. 

\citet{BP2005} have proposed the three visibility correlation to measure the EoR 21-cm BS from radio-interferometric data. In the present paper, we implement this to measure the BS using $\approx 17$ minutes of $154.2$ MHz MWA data with $(\mathrm{RA},\mathrm{DEC}) = (6.1^\circ, -26.7^\circ)$. The implementation is carried out in stages, which we briefly outline below.  \cite{Gill_2024_2d3vc} (hereafter \citetalias{Gill_2024_2d3vc}) proposed and validated an efficient technique to compute the three visibility correlation and estimate the angular BS (ABS) $B_A(\ell_1,\ell_2,\ell_3)$ of the sky signal from radio-interferometric data at a single frequency $\nu$.   \cite{Gill_2025_mabs} (hereafter \citetalias{Gill_2025_mabs}) generalized this for multi-frequency data to estimate the multi-frequency angular BS (MABS) $B_A(\ell_1,\ell_2,\ell_3,\nu_1,\nu_2,\nu_3)$. This, in turn, is used to estimate the cylindrical BS $B(k_{1\perp},k_{2\perp},k_{3\perp},k_{1\parallel},k_{2\parallel})$ and then the spherical BS  $B(k_1,k_2,k_3)$  \citep{gill_2024} of the three dimensional (3D) 21-cm signal. 
 
 It is now well accepted that foreground contamination in the estimated EoR 21-cm cylindrical PS $P(k_{1\perp},k_{1\parallel})$  is largely confined within the ``foreground wedge'' \citep{adatta10,parsons12,Morales_2012, Trott_2012ApJ,Vedantham_2012, Murray_2018}. The region of the $(k_{1\perp},k_{1\parallel})$ plane outside the foreground wedge, the ``EoR window'',  is relatively free of foreground contamination and can be used to measure the EoR 21-cm PS. In this work,  we provide a detailed analysis of the cylindrical BS to elucidate the impact of foreground contamination. In particular, we investigate whether it is possible to identify a foreground wedge and an EoR window for the 21-cm cylindrical BS. 
 
 In this paper, we present a comprehensive analysis of the spherical BS considering all possible triangle configurations that arise from the present observational data.  We use this to place upper limits on the mean-cubed 21-cm brightness temperature fluctuations, and compare these with the corresponding upper limits on the mean-squared brightness temperature fluctuations obtained from the same data.

The structure of the paper is as follows. In Section \ref{sec:data}, we describe the observational data and the simulation of system noise. Section \ref{sec:meth} provides a brief overview of the BS statistic, including its estimation and parameterization. The main results are presented in Section \ref{sec:results}. Finally, Section \ref{sec:sum} summarizes the findings and concludes the paper.

The cosmological parameters used in this work are taken from \citet{planck2020}.

\section{Data}
\label{sec:data}

We used data from the Phase II MWA drift-scan observations (Project ID: G0031) described in \citet{Patwa_2021}. These observations were made at a fixed declination of $\mathrm{DEC} = -26.7^\circ$, corresponding to the zenith, and span a range of right ascensions (RA) from $349^\circ$ to $70.3^\circ$, covering a total of $81.3^\circ$.
The data was recorded for 162 discrete pointing centers, spaced at an interval of $0.5^\circ$ along the RA. In this work, we restrict our analysis to a single pointing center corresponding to $ { \rm (RA, DEC)} =(6.1^\circ,-26.7^\circ)$, which was observed for a total of $\approx 17$ minutes across  $10$ nights.

The observation comprises $N_c = 768$ frequency channels, with frequency resolution of $\Delta\nu_c = 40$ kHz, yielding a total bandwidth of $B_{\rm bw} = 30.72$ MHz. The central frequency $\nu_c = 154.2$ MHz corresponds to \HI 21-cm line from  redshift  $z = 8.2$. The full bandwidth is divided into $24$ coarse bands, each containing $32$ fine channels. For each coarse band, four edge channels at both ends and one central channel are flagged. The resulting flagging pattern is periodic with a frequency spacing of $1.28$ MHz. Additionally, at each baseline, different channels may be flagged to avoid radio frequency interference (RFI). We have validated our 3D BS estimator on this data in \citetalias{Gill_2025_mabs}, using the same baseline distribution, frequency bandwidth, and flagging scheme. For each baseline $\u$ and frequency $\nu$, we have separate measurements for two orthogonal polarizations, namely LL and RR, respectively,  denoted  $\V_{\rm LL}(\u, \nu)$ and $\V_{\rm RR}(\u, \nu)$. We treat these as independent measurements and combine them when gridding the visibility data. 

We note that the same data was used in \citet{chatterjee_2024} and \citet{Elahi_missing}  to estimate the MAPS $C_{\ell}(\Delta \nu)$ and the cylindrical PS $P(k_{1\perp},k_{1\parallel})$ considering an approach that is very similar to the one adopted here to estimate the MABS and the cylindrical BS. Furthermore, the baseline distribution, frequency coverage, and flagging pattern of the simulated data used to validate our  BS estimators in \citetalias{Gill_2024_2d3vc} and \citetalias{Gill_2025_mabs} exactly match those of the data that has been analyzed here.

To characterize the system noise contribution to the BS, we perform dedicated `noise-only' simulations. The noise in the visibilities is modeled as Gaussian random variables with zero mean and standard deviation of $\sigma = 20~\mathrm{Jy}$, calculated using the analytical expression given in Eq. (1) of \citet{chatterjee_2024}. We assume the noise to be uncorrelated across different baselines, frequency channels, timestamps, and polarizations. Simulated visibilities are generated using the same baseline and frequency configurations as the actual data and incorporating identical flagging information. To ensure statistical robustness, we produce 50 independent noise realizations. These simulations are used to quantify the uncertainty in the measured BS due to system noise.

\section{Methodology}
\label{sec:meth}

We employ our recently developed estimator (\citetalias{Gill_2025_mabs}) to compute the BS from $\V(\u,\nu)$ the radio interferometric visibility data. While the detailed formulation and validation of the estimator can be found in \citetalias{Gill_2025_mabs}, a concise overview of the procedure is presented here. The estimation proceeds in three main steps. In the first step, we evaluate the multi-frequency angular BS (MABS), $B_A(\ell_1, \ell_2, \ell_3, \nu_1, \nu_2, \nu_3)$, by correlating three gridded visibilities across different baselines $(\u_1, \u_2, \u_3)$ and frequencies $(\nu_1, \nu_2, \nu_3)$. In the second step, we perform a 2D Fourier transform of the MABS to compute the 3D cylindrical BS. In the third step, we average triangles of different orientations to calculate the spherical BS.  
The following subsections detail the procedures and results for the first two steps. The results for the spherical BS are presented in the next section.

\subsection{The MABS}
The relation between the MABS and the three visibility correlation is given by,
\begin{equation}
\begin{aligned}
 B_A({\ell_1}, {\ell_2}&, {\ell_3},\nu_1,\nu_2,\nu_3) = \dfrac{3}{\pi\theta_0^2 Q^3} \exp\left [{\pi^2\theta_0^2\Delta U^2/3} \right ] \\& \times\langle \V(\u_1,\nu_1) \V(\u_2,\nu_2) \V(\u_3+\Delta \u,\nu_3)\rangle \,.
\label{eq:mabs}
\end{aligned}
\end{equation}
where $\u_1 + \u_2 + \u_3=0$ forms a closed triangle, where $\bfl=2\pi \u$ and $\Delta \u$ is the deviation from a closed triangle configuration. Here $Q=2 k_B/\lambda^2$ and $\theta_0=0.6~\theta_{\rm FWHM}$, where $\theta_{\rm FWHM}\approx24.68^\circ$ for the MWA primary beam (PB). 
 We use the values of $Q$ and $\theta_0$ at the central frequency $\nu_c$, assuming the bandwidth is sufficiently small. 

\citetalias{Gill_2025_mabs} presents the binned MABS estimator, and validates it using simulated visibility data that exactly match the observation for which we now analyze the actual data. The procedure here is exactly identical to \citetalias{Gill_2025_mabs}, and we do not present it here again. Briefly, we grid the visibilities on a square grid of spacing $\Delta U_g = \sqrt{\ln 2} / (\pi \theta_0) \approx 1$ in the $\u$-plane.
We then divide the $\u$ plane into several concentric annular rings. Each combination of three annular rings provides an estimate of the binned MABS.   We have evaluated the binned MABS estimator for all possible combinations of three rings. Note that some combinations of three rings do not contain any closed triangle $\U_1+\U_2+\U_3=0$ and are discarded.  

\begin{figure*}
\centering
\includegraphics[width=1\textwidth]{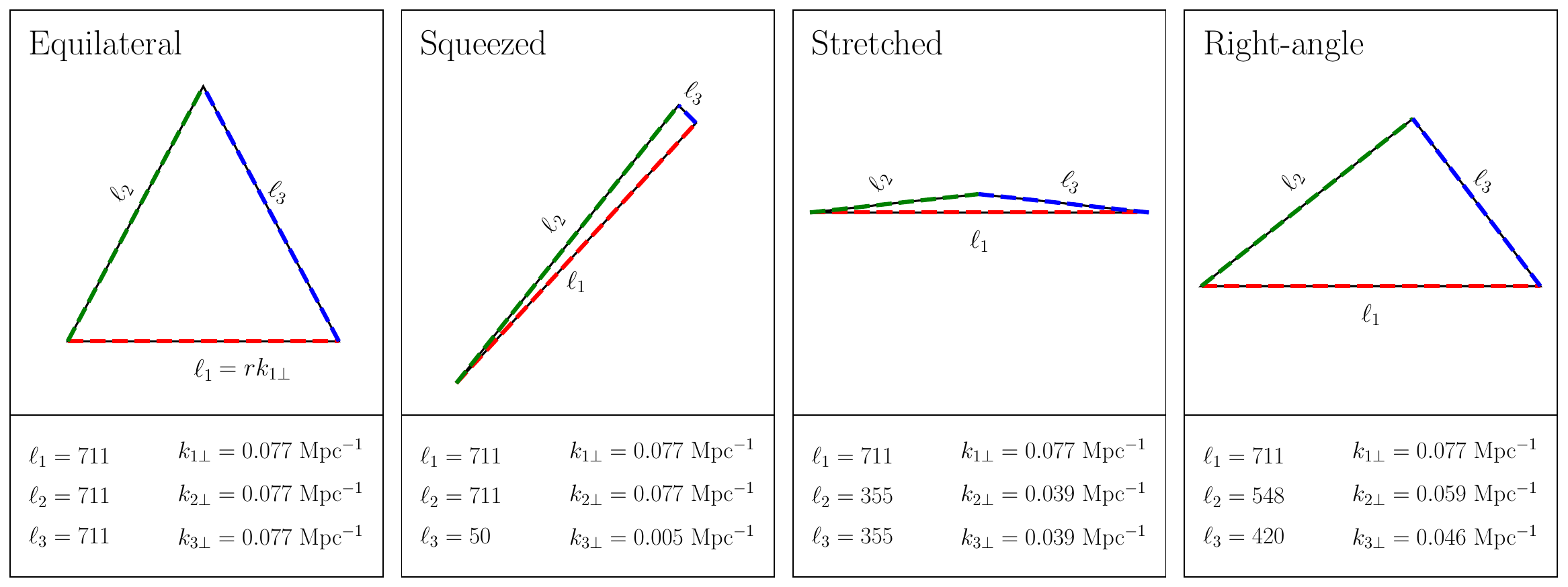}
\caption{Four triangle configurations—equilateral, squeezed, stretched, and right-angle—for which we explicitly present the MABS and the 3D cylindrical BS results. For each case, the numerical values of the triangle sides ($\ell_1,\ell_2,\ell_3$) together with the corresponding perpendicular wavenumbers $k_{1\perp}=\ell_1/r$, $k_{2\perp}=\ell_2/r$, and $k_{3\perp}=\ell_3/r$ are listed in the bottom row. These four cases are a small subset of the total of $669$ distinct triangle configurations formed by all the $(\bfl_1,\bfl_2,\bfl_3)$ analyzed in this study.}
\label{fig:tri_confi}
\end{figure*}

The sky signal is assumed to be isotropic in the plane of the sky, 
which allows us to bin together triangles $(\U_1,\U_2,\U_3)$, or equivalently $(\bfl_1,\bfl_2,\bfl_3)$,  of the same shape and size but with different orientations. Here we obtain estimates of the MABS, $B_A(\ell_1, \ell_2, \ell_3, \nu_1, \nu_2,\nu_3)$ for 669 distinct $(\ell_1, \ell_2, \ell_3)$ bins that together cover all possible triangle configurations for the present data. In addition to triangles of different orientations,  each bin also includes triangles within a finite range of shapes and sizes, as determined by the widths of the annular rings. In the subsequent discussion, we refer to each bin, labeled $(\ell_1, \ell_2, \ell_3)$ with $\ell_1 \ge \ell_2 \ge \ell_3$, as a ``triangle configuration''. We use $\ell_1$, the largest side, to quantify the size of the triangle configuration,  and the sides span the range $22 \leq \ell_1, \ell_2,\ell_3  \leq 1629$.
Here we explicitly show the MABS for four representative triangle configurations, all of the same size  $(\ell_1=711$) but of different shapes, namely equilateral, squeezed, stretched, and right-angled, shown from left to right in the top row of Figure \ref{fig:tri_confi}. The corresponding 
$(\ell_1, \ell_2, \ell_3)$ values are shown in the bottom row of the figure.

\begin{figure*}
\centering
\includegraphics[width=1\textwidth]{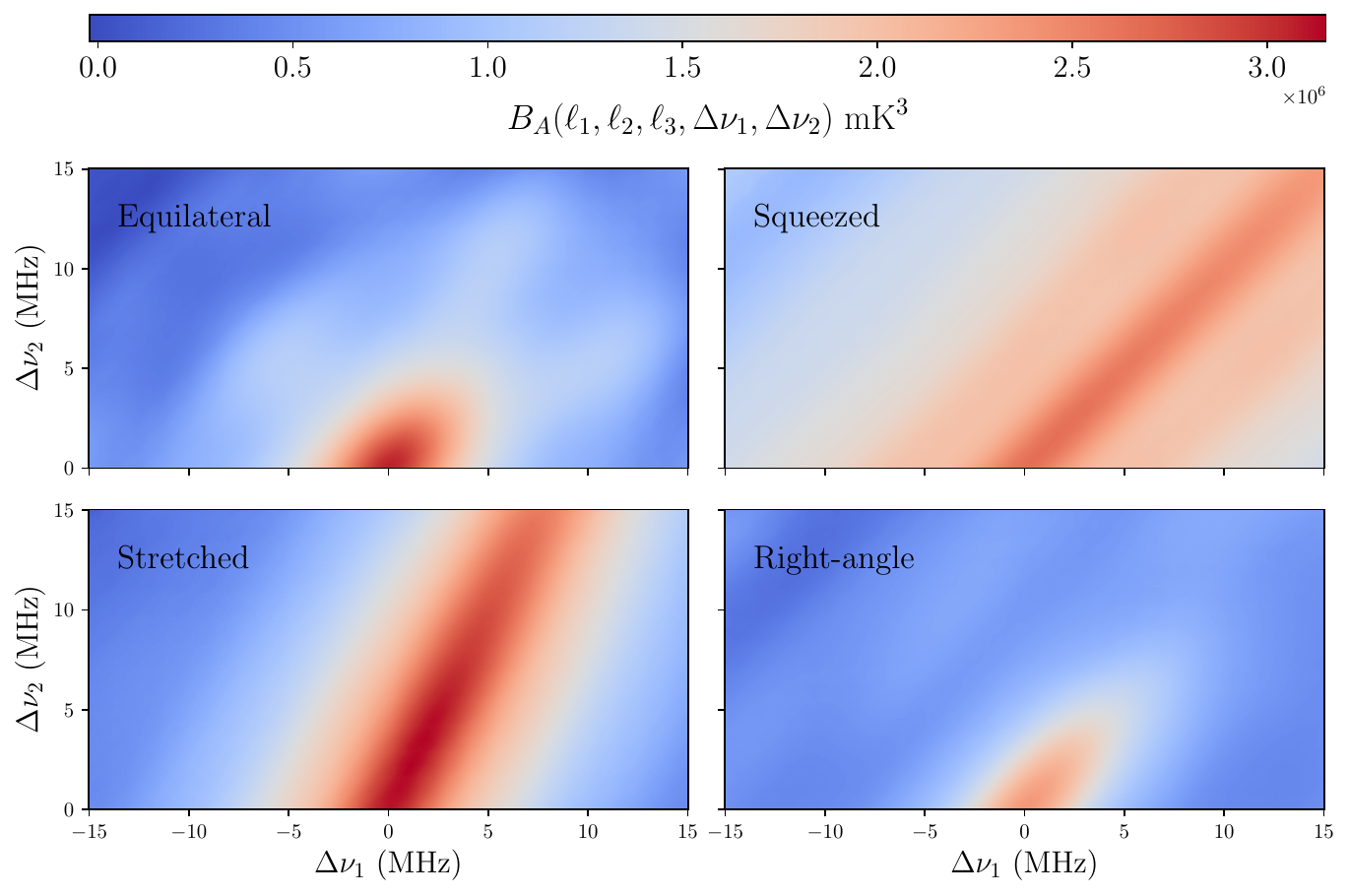}
\caption{MABS $B_A(\ell_1,\ell_2,\ell_3,\Delta\nu_1,\Delta\nu_2)$ plotted as a function of $(\Delta\nu_1,\Delta\nu_2)$ for the four representative triangle configurations— equilateral (top-left), squeezed (top-right), stretched (bottom-left), and right‐angle (bottom-right)—illustrated in Figure~\ref{fig:tri_confi}. The $\Delta\nu_1-\Delta\nu_2$ plane is restricted to $|\nu_1|,|\nu_2|\leq B_{\rm bw}/2$, thereby excluding the frequency separations that are not sampled. Only the upper half-plane, i.e. $B_{\mathrm{bw}}/2 \le \Delta\nu_1 \le B_{\mathrm{bw}}/2$ and $0 \le \Delta\nu_2 \le B_{\mathrm{bw}}/2$, is displayed, owing to the symmetry
$B_A(\ell_1,\ell_2,\ell_3,-\Delta\nu_1,-\Delta\nu_2)=B_A(\ell_1,\ell_2,\ell_3,\Delta\nu_1,\Delta\nu_2)$.}
\label{fig:mabs}
\end{figure*}

Considering the MABS $B_A(\ell_1, \ell_2, \ell_3, \nu_1, \nu_2,\nu_3)$, we assume that the 21-cm signal is ergodic along frequency, or equivalently, the line of sight (LoS) direction. As a result, the MABS depends only on the frequency separations, i.e. we can then express the MABS as $B_A(\ell_1, \ell_2, \ell_3, \Delta\nu_1, \Delta\nu_2)$, where $\Delta\nu_1 = \nu_1 - \nu_3$ and $\Delta\nu_2 = \nu_2 - \nu_3$.  For any triangle configuration $(\ell_1, \ell_2, \ell_3)$, we have discrete estimates of the MABS 
at a regular interval of $\Delta\nu_c = 40~\mathrm{kHz}$ over the range $-B_{\rm bw}$ to $ B_{\rm bw}$ in the $(\Delta\nu_1, \Delta\nu_2)$ plane. However, the region $|\Delta\nu_2 - \Delta\nu_1| > B_{\rm bw}$ remains unsampled (Figure~2 of \citetalias{Gill_2025_mabs}),  and we  restrict the analysis  to the range $-B_{\rm bw}/2$ to $ B_{\rm bw}/2$ for the ease of computation. Furthermore, the MABS exhibits the  symmetry $B_A(\ell_1, \ell_2, \ell_3, -\Delta\nu_1, -\Delta\nu_2)=B_A(\ell_1, \ell_2, \ell_3, \Delta\nu_1, \Delta\nu_2)$, and it suffices to consider only the upper half-plane $-B_{\rm bw}/2 \le \Delta \nu_1 \le B_{\rm bw}/2$ and $0 \le \Delta \nu_2 \le B_{\rm bw}/2$, as in Figure \ref{fig:mabs}.

\begin{figure}
\centering
\includegraphics[width=.5\textwidth]{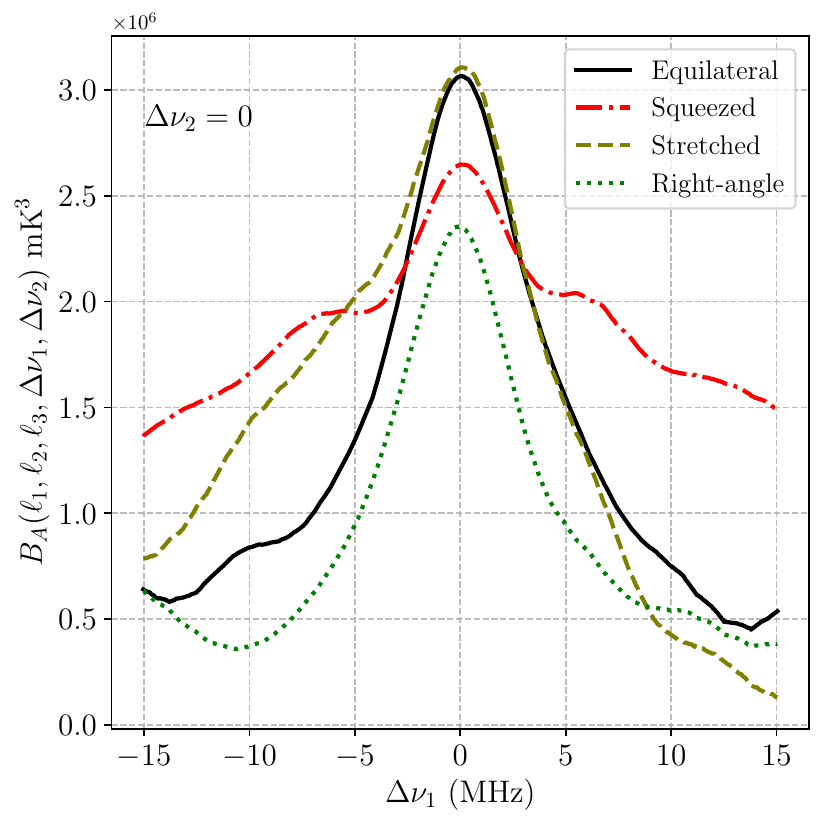}
\caption{One-dimensional slices of the MABS plotted as a function of $\Delta\nu_1$ at fixed $\Delta\nu_2 = 0$, corresponding to each of the 2D MABS panels presented in Figure \ref{fig:mabs}. }
\label{fig:mabs_slice}
\end{figure}

Figure~\ref{fig:mabs} presents the MABS, $B_A(\ell_1, \ell_2, \ell_3, \Delta\nu_1, \Delta\nu_2)$, as a function of $(\Delta\nu_1, \Delta\nu_2)$. The four panels of this figure respectively correspond to the four $(\ell_1, \ell_2, \ell_3)$
triangle configurations shown in Figure.~\ref{fig:tri_confi}. We have estimates of the MABS across the entire $(\Delta\nu_1, \Delta\nu_2)$ range shown here, and there are no missing frequency separations $(\Delta\nu_1, \Delta\nu_2)$ due to the missing frequency channels $\nu$ of the MWA data. However, the number of triplets $(\nu_1,\nu_2,\nu_3)$ corresponding to each combination of $(\Delta\nu_1, \Delta\nu_2)$ exhibits a modulation due to the missing frequency channels, as shown in Figure~2 of \citetalias{Gill_2025_mabs}. The sky signal here is foreground-dominated, and we may interpret the estimated MABS as arising from the foregrounds.  In all cases, we see that the MABS attains its maximum value $(\sim 10^6 \, {\rm mK}^3)$ at $\Delta\nu_1 = \Delta\nu_2 = 0$ and slowly declines with increasing frequency separations.  The decline is not isotropic in $(\Delta\nu_1, \Delta\nu_2)$, and is faster in the second quadrant compared to the first. The exact nature of the decline varies with the triangle shape, and we see that for the squeezed and stretched triangles, the red region (large value of the MABS) covers a wider range of $(\Delta\nu_1, \Delta\nu_2)$ compared to the equilateral and right-angle triangles. 

Figure~\ref{fig:mabs_slice} shows the MABS, $B_A(\ell_1, \ell_2, \ell_3, \Delta\nu_1, \Delta\nu_2)$, as a function of $\Delta\nu_1$ at fixed $\Delta\nu_{2}=0$. The different line styles correspond to different triangle configurations, as indicated in the figure legend.  The maximum value of the MABS, at $\Delta\nu_1= \Delta\nu_2=0$, varies in the range $3.1\times10^6 - 2.4\times10^6 \, {\rm mK}^3$, depending on the triangle configuration. 
In all cases, the MABS declines smoothly with increasing $\mid \Delta \nu_1 \mid $ for $\mid \Delta \nu_1 \mid  \le 5 \, {\rm MHz}$, beyond which it continues to decline, but also exhibits some structures. This is particularly noticeable for the squeezed triangles, where the decline flattens out and shows small oscillations at $\mid \Delta \nu_1 \mid  \ge 5 \, {\rm MHz}$.  
In no case does the MABS completely decorrelate (approach zero) within the $\mid \Delta \nu_1 \mid \le 15.36 \, {\rm MHz}$ range considered here. This behavior reflects the smooth spectral nature of the foregrounds, while the oscillations and other complex features that appear in the MABS are possibly due to additional chromaticity introduced by instrumental effects.   Note the contrast with respect to the simulated 21-cm signal where the MABS decorrelates to $95\%$ of its peak value at $\mid \Delta\nu_1 \mid  = 2$ MHz, and is close to zero at $\mid \Delta\nu_1 \mid  \ge  2 \, {\rm MHz}$ (Figure 3 of \citetalias{Gill_2025_mabs}).

\subsection{The Cylindrical BS}

The 3D cylindrical BS is obtained by performing a 2D Fourier transform of the MABS, given by \citep{Bharadwaj2005},
\begin{equation}
    \begin{aligned}
 &B(k_{1 \perp},k_{2 \perp},k_{3 \perp},k_{1\parallel},k_{2\parallel}) = 
 r^4 r'^2 \int d(\Delta \nu_1) \int d(\Delta \nu_2)  
\\&\times \exp{[-ir'(k_{1 \parallel}\Delta \nu_1+k_{2 \parallel}\Delta \nu_2)]} \, B_A(\ell_1, \ell_2, \ell_3, \Delta \nu_1, \Delta \nu_2)~,
\end{aligned}
\label{eq:3dbs_mabs}
\end{equation}
where $r$ is the comoving distance corresponding to the 21-cm signal received at $\nu_c$, which can be calculated from the redshift $z=\left(\frac{1420 ~\rm{MHz}}{\nu}\right)-1$, and $r^\prime=\frac{dr}{d\nu}$. Considering the wave number $\kk$, here $\kp={\bfl}/{r}$ and $\kll$  are respectively the components of $\kk$  perpendicular and parallel to the LoS direction. The cylindrical BS $B(k_{1 \perp},k_{2 \perp},k_{3 \perp},k_{1\parallel},k_{2\parallel})$ is a function of five independent variables,  and $k_{3\parallel}$ is determined by the closed triangle condition $k_{3\parallel}=-k_{1\parallel}-k_{2\parallel}$. 
 The 2D Fourier transform of the MABS with respect to $( \Delta \nu_1, \Delta \nu_2)$ (Eq.~\ref{eq:3dbs_mabs}) 
 introduces ripples along  $(k_{1\parallel},k_{2\parallel})$  in the 3D cylindrical BS due to the abrupt truncation at the boundaries $ \pm 15.36 \, {\rm MHz}$.  We introduce a 2D Blackman–Nuttall window function \citep{Nuttall_1981} that smoothly tapers the MABS at the boundaries, which suppresses the edge effects and mitigates the ripples along $(k_{1\parallel},k_{2\parallel})$.

\begin{figure*}
\centering
\includegraphics[width=1\textwidth]{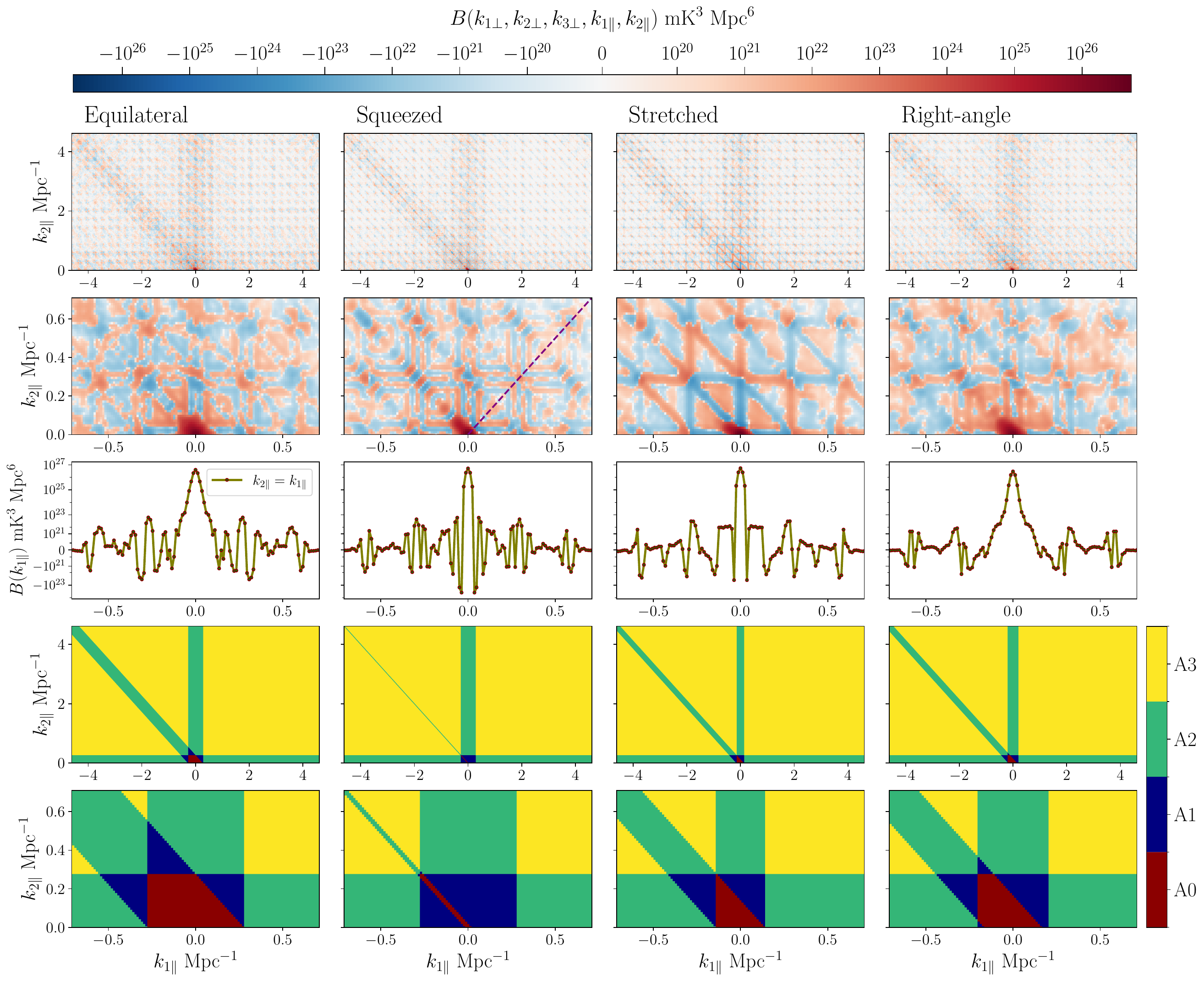}
\caption{The 3D cylindrical BS $B(k_{1\perp}, k_{2\perp}, k_{3\perp}, k_{1\parallel}, k_{2\parallel})$  for the four triangle configurations ($k_{1\perp},k_{2\perp},k_{3\perp}$) shown in Figure~\ref{fig:tri_confi}. These results are obtained via a 2D Fourier transform (Eq.~\ref{eq:3dbs_mabs}) of the MABS shown in Figure~\ref{fig:mabs}. The top row shows heat-maps of the 3D BS over the entire $(k_{1\parallel}, k_{2\parallel})$ plane, while the second row zooms into the central region $-0.7~{\rm Mpc}^{-1}\leq(k_{1\parallel}, k_{2\parallel})\leq0.7~{\rm Mpc}^{-1}$. The third row shows 1D cuts along the diagonal $k_{1\parallel}=k_{2\parallel}$ (marked by a dashed brown line in the second row), plotted on a symmetric logarithmic scale with a linear range of $\pm 10^{21}~{\rm mK^3~Mpc^6}$. The fourth row shows four distinct regions (A0, A1, A2, A3) into which the $(k_{1\parallel}, k_{2\parallel})$ plane has been divided based on the number of triangle sides that are 
 outside the foreground wedge, that is,   $(k_{i\parallel}  > [r/(r'\Delta\nu_c)]k_{i_\perp})$. The bottom row provides a zoomed‐in view of the fourth row. }
\label{fig:cylbs}
\end{figure*}

Figure~\ref{fig:cylbs} shows the 3D cylindrical BS $B(k_{1 \perp},k_{2 \perp},k_{3 \perp},k_{1\parallel},k_{2\parallel})$ as a function of $(k_{1\parallel},k_{2\parallel})$ with $(k_{1 \perp},k_{2 \perp},k_{3 \perp})=r^{-1} \, (\ell_1, \ell_2, \ell_3)$ fixed. Each column corresponds to a particular triangle configuration in the $\kk_{\perp}$ plane, with the $(k_{1 \perp},k_{2 \perp},k_{3 \perp})$ values shown in the bottom row of Figure~\ref{fig:tri_confi}. 
The  available $(k_{1\parallel}, k_{2\parallel})$ values span the range $[-4.61~\rm{Mpc}^{-1}, 4.61~\rm{Mpc}^{-1}]$ with a uniform spacing $\Delta k_{\parallel}=0.012~\rm{Mpc}^{-1}$. The cylindrical BS has the symmetry $B(k_{1 \perp},k_{2 \perp},k_{3 \perp},k_{1\parallel},k_{2\parallel})=B(k_{1 \perp},k_{2 \perp},k_{3 \perp},-k_{1\parallel},-k_{2\parallel})$ and it suffices to show the results for only the upper half plane, as in Figure~\ref{fig:cylbs}. 

The top row of Figure~\ref{fig:cylbs} shows the cylindrical BS obtained from the MABS shown in Figure~\ref{fig:mabs}. The panels here consider the entire available $(k_{1\parallel}, k_{2\parallel})$  range.  We see that the cylindrical BS peaks at $k_{1\parallel}=k_{2\parallel}=0$ and falls rapidly at larger values of $(k_{1\parallel},k_{2\parallel})$. In addition to the central peak, we also notice three dark bands that are respectively aligned with $k_{1\parallel} = 0$, $k_{2\parallel} = 0$, and  $k_{1\parallel}=-k_{2\parallel}$  ($k_{3\parallel} = 0$). We identify these as the signature of spectrally smooth foreground contamination, which we will explore in greater detail in section~\ref{sec:fgw}.  
In addition, we see a periodic square grid pattern that runs across the entire $(k_{1\parallel}, k_{2\parallel})$ plane. This has a spacing of $\delta k_\parallel = 0.29~\mathrm{Mpc}^{-1}$, which arises from the periodic flagging present in the MWA data. A similar effect is also observed in the 3D cylindrical PS and has been studied in detail by \citet{Elahi_missing}.

The second row provides a zoomed-in view of the top row limited to the range $-0.7~{\rm Mpc}^{-1} \leq (k_{1\parallel}, k_{2\parallel}) \leq 0.7~{\rm Mpc}^{-1}$, where the BS signal is concentrated. We see that overall, the BS has large values in this region compared to the large $(k_{1\parallel}, k_{2\parallel})$ that have been excluded from these panels. A prominent feature across all configurations is the bright central region, which is elongated, roughly along the anti-diagonal direction.  The direction of this elongation is orthogonal to that of the MABS shown in Figure~\ref{fig:mabs}. This is expected as the two are related through a Fourier transform (Eq.~\ref{eq:3dbs_mabs}). The central bright region of the cylindrical BS is strongly foreground dominated, and the values of the cylindrical BS  fall at larger  $k_{\parallel}$. However, we find considerable foreground leakage outside the central bright region. The leakage shows a complex periodic pattern of positive and negative values, which is a combination of the spectral variation of the foregrounds, the telescope's chromatic response, and the periodic pattern of missing frequency channels.

The third row shows the cylindrical BS $B(k_{1 \perp},k_{2 \perp},k_{3 \perp},k_{1\parallel},k_{2\parallel})$ as a function of $k_{1\parallel}$ along the diagonal $k_{1\parallel} = k_{2\parallel}$, which is indicated by the dashed line in the panel for the squeezed triangle in the second row. In all cases, the peak value of the cylindrical BS  is approximately $10^{27}~{\rm mK}^3~{\rm Mpc}^6$, and it decreases rapidly with increasing $k_{1\parallel}$, and subsequently exhibits an oscillatory behavior. The onset of these oscillations depends on the triangle configuration. For the equilateral triangle, the 
oscillations begin around $k_{1\parallel} \approx 0.1~{\rm Mpc}^{-1}$, while it begins earlier $(\approx 0.03~{\rm Mpc}^{-1})$ for the squeezed and stretched triangles, and later at  $k_{1\parallel} \approx 0.2~{\rm Mpc}^{-1}$  for the right-angle triangle. 
The amplitude of these oscillations is in the range of $10^{21}$–$10^{22}~{\rm mK}^3\,{\rm Mpc}^6$, making them $5$–$6$ orders of magnitude smaller than the maximum value of the BS. 
We find values in the range $\pm (10^{19} -10^{20}) \,  {\rm mK}^3\,{\rm Mpc}^6$ in the region where the cylindrical BS appears to have values close to zero in the figure. We have analyzed the noise-only simulations in exactly the same way as the actual data, and used the r.m.s. of the BS from $50$ realizations to estimate $\delta B_N$ which quantifies the statistical fluctuations that we expect from the system noise. We find that the estimated values of the cylindrical BS are considerably in excess of $\delta B_N$, indicating that all our results are foreground dominated.

\subsection{The Foreground Wedge}
\label{sec:fgw}
The foreground contamination in the cylindrical PS $P(k_{1\perp},k_{1\parallel})$ is largely confined within a wedge in the  $(k_{1\perp},k_{1\parallel})$ plane  \citep{adatta10,parsons12,Morales_2012, Vedantham_2012, Murray_2018}. 
The wedge is defined by the condition $k_{1\parallel} \leq  [k_{1\parallel}]_H$, where the wedge boundary 
\begin{equation}
 [k_{1\parallel}]_H= \left[\frac{r}{r'\Delta\nu_c}\right] k_{1\perp}=3.51 k_{1\perp}
 \label{eq:wb}
\end{equation}
corresponds to a source at the horizon, and we have the value $3.51$ for the data considered here. The region outside the wedge, the EoR window, is relatively free of foreground contamination. In principle, it is possible to overcome the problem of foreground contamination by adopting the technique of ``Foreground Avoidance'', which avoids the $(k_{1\perp},k_{1\parallel})$ modes within the 
foreground wedge and only uses those in the EoR window to estimate the 21-cm PS.   

The foreground wedge is a straightforward consequence of baseline migration, and we also expect this to manifest itself in the cylindrical BS $B(k_{1 \perp},k_{2 \perp},k_{3 \perp},k_{1\parallel},k_{2\parallel})$ as well. However, the difficulty lies in visualizing this, the cylindrical BS being a function of five variables. Considering a fixed triangle configuration $(k_{1 \perp},k_{2 \perp},k_{3 \perp})$, we have three distinct values for the wedge boundary  
\begin{equation}
     \left([k_{1\parallel}]_H,  [k_{2\parallel}]_H, [k_{3\parallel}]_H\right) = 
\left[\frac{r}{r'\Delta\nu_c}\right] \times (k_{1 \perp},k_{2 \perp},k_{3 \perp})
\label{eq:bwb}
\end{equation}
and we have three conditions 
\begin{equation}
    \mid k_{a \parallel}\mid  \le   [k_{a \parallel}]_H
    \label{eq:wedge}
\end{equation}
 for $a=1,2$ and $3$ respectively. 

The fourth row of Figure~\ref{fig:cylbs} shows the region of $(k_{1\parallel},k_{2\parallel})$ space that is denoted by Eq.~(\ref {eq:wedge}) for the four different triangle configurations considered here.  In each panel, the vertical and horizontal green bands correspond to $-[k_{1 \parallel}]_H \le k_{1 \parallel} \le [k_{1 \parallel}]_H$ and 
$0 \le k_{2 \parallel} \le [k_{2 \parallel}]_H$ respectively. The anti-diagonal 
green band is centered around $k_{3 \parallel}=0$ i.e. $k_{2 \parallel}=-k_{1 \parallel}$, and it has a width of $2 \times [k_{3 \parallel}]_H$. 
We expect the foreground contamination to be largely confined within the region covered by the green bands, and the yellow region is expected to be relatively free of foreground contamination. Here, we may interpret the yellow region as the EoR window. Considering the top row of Figure~\ref{fig:cylbs}, we see the bands corresponding to $k_{1 \parallel}=0$, $k_{2 \parallel}=0$ and $k_{3 \parallel}=0$ are clearly visible for the actual data, and these regions show a high level of foreground contamination relative to the rest of the $(k_{1\parallel},k_{2\parallel})$ plane. We notice that the EoR window is not as clean as expected, and we find considerable foreground leakage due to the periodic pattern of missing frequency channels.

The bottom row of Figure~\ref{fig:cylbs} provides a zoomed-in view of the central region, within the range $-0.7~{\rm Mpc}^{-1} \leq (k_{1\parallel}, k_{2\parallel}) \leq 0.7~{\rm Mpc}^{-1}$ that matches the second row.  
Here we can clearly identify four distinct regions in the $(k_{1\parallel},k_{2\parallel})$ plane, which we respectively label 
A0, A1, A2, and A3, as indicated in the figure. Here, A3 refers to the region where all three modes $(k_{1 \perp},k_{1 \parallel})$, $(k_{2 \perp},k_{2 \parallel})$ and $(k_{3 \perp},k_{3 \parallel})$ avoid the foreground wedge. Similarly, A2, A1, and A0 refer to the regions where, respectively, only two, one, and zero modes avoid the foreground wedge. We expect foreground contamination to be most severe in region A0, which corresponds to the bright central region in the panels of the second row. We expect the foreground contamination to diminish as we go from A0 to A3. As mentioned earlier, we refer to A3 as the EoR window for the cylindrical BS.  Considering the fourth row, we see that the EoR window (A3) covers the largest fraction of the $(k_{1\parallel},k_{2\parallel})$ plane, and the fraction shrinks successively as we move from A3 to A0.  Table~\ref{tab:ntri_by_region_horizontal} shows the number of triangles corresponding to each of the four regions considered here. We see that $83.6 \%$ of the triangles are in the EoR window (A3), which is very promising for detecting the EoR 21-cm BS. Considering foreground avoidance, we need to discard only a small fraction of the triangles to avoid foreground contamination. Note that the values quoted in Table~\ref{tab:ntri_by_region_horizontal} refer to the total number of triangles used in our analysis, which is also the total number of triangles ($\kk_1,\kk_2,\kk_3$)  available from the MWA data, and not just those shown in Figure~\ref{fig:cylbs}.

\begin{table}[htbp]
  \centering
  \caption{Number of triangles categorized by the level of foreground contamination. The A0, A1, A2, and A3 correspond to regions where 0, 1, 2, and 3 sides of a triangle avoid the foreground-dominated zone, respectively.  The total number of triangles analyzed is $1.66 \times 10^{12}$.}
  \label{tab:ntri_by_region_horizontal}
  \begin{tabular}{@{} l c c c c @{}}  
    \toprule
    \textbf{Region} & \textbf{A0} & \textbf{A1} & \textbf{A2} & \textbf{A3} \\
    \midrule
    \textbf{\(N_{\rm tri}\)}      & \(2.45\times10^{9}\)   & \(5.75\times10^{9}\)   & \(2.65\times10^{11}\) & \(1.39\times10^{12}\)  \\
    Percentage   & 0.15\%            & 0.35\%            & 15.91\%            & 83.60\%              \\
    \bottomrule
  \end{tabular}
\end{table}


\begin{figure*}
\centering
\includegraphics[width=1\textwidth]{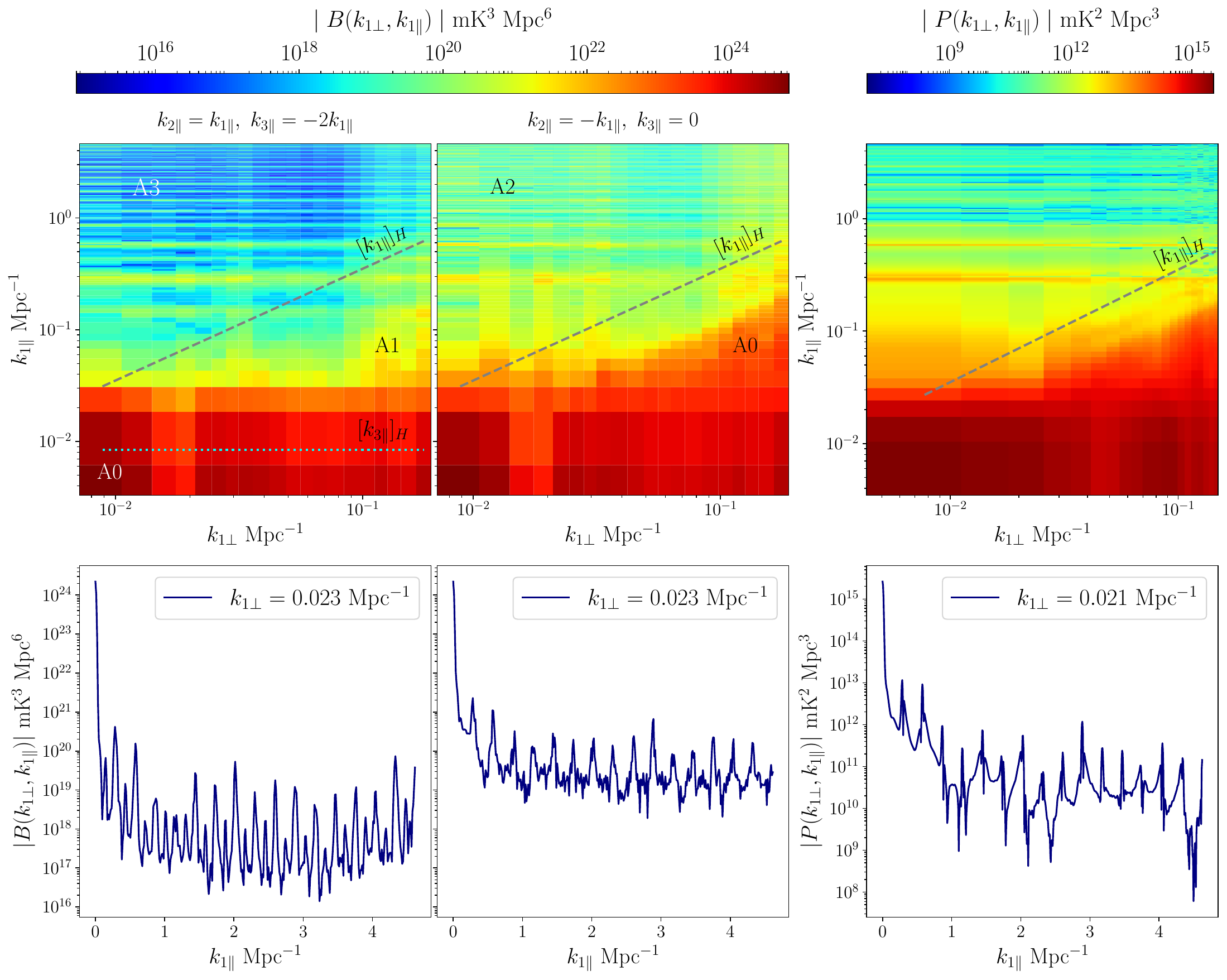}
\caption{The top row shows $\mid B(k_{1\perp},k_{1\parallel}) \mid$, which is  the magnitude of the 3D cylindrical BS $|B(k_{1\perp}, k_{2\perp}, k_{3\perp}, k_{1\parallel}, k_{2\parallel})|$ for squeezed triangles $(k_{1\perp}\approx k_{2\perp}, \,k_{3\perp}\rightarrow0)$ with the constraints $k_{1\parallel}=k_{2\parallel}$ and $k_{1\parallel}=- k_{2\parallel}$ for the left and center panels respectively.  The top right panel shows $\mid P(k_{1\perp},k_{1\parallel}) \mid$ the cylindrical PS for the same data from \citep{Elahi_missing}. In each panel, the dashed lines shows the predicted foreground wedge boundary   $[k_{1\parallel}]_H  = [r/(r'\Delta\nu_c)]k_{1\perp}$, whereas the dotted line in the left panel shows $[k_{3\parallel}]_H$   corresponding to the minimum value of $ k_{3\perp}$  $(=0.002 \, {\rm Mpc}^{-1})$.   The panels in the bottom row  show 1D sections through the respective panels of the upper row,  at  $k_{1\perp} \approx 0.023~\mathrm{Mpc}^{-1}$ for the BS and $k_{1\perp} \approx 0.021~\mathrm{Mpc}^{-1}$ for the PS.}
\label{fig:wedge_sq}
\end{figure*}

Figure~\ref{fig:wedge_sq} shows a comparison of $B(k_{1\perp}, k_{1\parallel})$ 
the cylindrical BS  with $P(k_{1\perp}, k_{1\parallel})$  the cylindrical PS estimates from \citet{Elahi_missing} who have analyzed the same MWA data.   The first two columns show
$B(k_{1\perp}, k_{2\perp}, k_{3\perp}, k_{1\parallel}, k_{2\parallel})$
as a function of $(k_{1\perp}, k_{1\parallel})$ considering squeezed triangles for which $k_{2\perp}\approx k_{1\perp}$ and $k_{3\perp}\rightarrow 0$  (actually  the minimum value $k_{3\perp}=0.002 \, {\rm Mpc}^{-1} $). The left and middle columns, respectively, consider ($k_{2\parallel} = k_{1\parallel}$, $k_{3\parallel} = -2k_{1\parallel}$) and ($k_{2\parallel} =- k_{1\parallel}$, $k_{3\parallel} = 0$).  The right column shows $P(k_{1\perp}, k_{1\parallel})$ .

We first consider the upper row of Figure~\ref{fig:wedge_sq}, which  shows heat maps of $B(k_{1\perp}, k_{1\parallel})$ and $P(k_{1\perp}, k_{1\parallel})$. The foreground wedge is clearly visible in the right panel that shows $P(k_{1\perp}, k_{1\parallel})$. We also notice leakage beyond the wedge boundary $(k_{1\parallel} \ge [k_{1\parallel}]_H)$, particularly horizontal streaks along $k_{1\parallel}$ due to the periodic pattern of missing frequency channels. 
The panels showing $B(k_{1\perp}, k_{1\parallel})$ both exhibit a foreground wedge similar to the right panel; however, there are differences in details. 
In the left panel, we have a large A3 region $(k_{1\parallel} > [k_{1\parallel}]_H)$ that has a relatively low level of foreground contamination. The region $[k_{1\parallel}]_H \ge k_{1\parallel} > [k_{3\parallel}]_H$ is A1, and the level of foreground contamination here is higher than that of the A3 region.  The region $k_{1\parallel} \le [k_{3\parallel}]_H$ approximately denotes the boundary of the A0 region where we have the highest level of foreground contamination. In the middle panel, we have $k_{3\parallel} = 0$ throughout, which is always inside the foreground wedge. Here, the region $k_{1\parallel} > [k_{1\parallel}]_H$ is A2, for which the level of foreground contamination is larger compared to the A3 region in the left panel. 
The region $ k_{1\parallel} \le [k_{1\parallel}]_H$ is A0, and we have a high level of foreground contamination here.  Furthermore, we observe a periodic pattern of horizontal streaks that extend across the $(k_{1 \perp}, k_{1 \parallel})$ plane. These streaks arise from the regular pattern of missing frequency channels in the data, similar to the cylindrical PS.

The panels in the bottom row of Figure~\ref{fig:wedge_sq} show a section through the corresponding panels of the upper row. The right panel shows $\mid P(k_{1\perp},k_{1\parallel})\mid $ as a function of $k_{1\parallel}$ at a fixed value of $k_{1\perp}=  0.021~\rm{Mpc}^{-1}$. Note the periodic pattern of spikes that occur at an interval $\delta k_{1\parallel} = 0.29~\rm{Mpc}^{-1}$ that exactly matches the period $($1.28$ \, {\rm MHz})$ of the flagged frequency channels. The middle and left panels show $\mid B(k_{1\perp}, k_{1\parallel})\mid$ as a function of $k_{1\parallel}$  for a fixed value of $k_{1 \perp}=  0.023~\rm{Mpc}^{-1}$. The middle panel shows a periodic pattern of spikes, whose period 
$\delta k_{1 \parallel} = 0.29~\rm{Mpc}^{-1}$ exactly matches the PS. In contrast, the left panel shows spikes with a regular spacing of $\delta k_{1 \parallel} = 0.145~\rm{Mpc}^{-1}$, which is half that of the other two panels. Furthermore, the amplitude of the spikes varies periodically with a period $\delta k_{1 \parallel} = 0.29~\rm{Mpc}^{-1}$, i.e., every alternate spike is large. The pattern seen in the left panel actually arises from a combination of two sets of spikes, respectively, corresponding to $k_{1 \parallel}$ and  $|k_{3\parallel}| = 2|k_{1\parallel}|$. The findings here substantiate our interpretation of the top row of  Figure~\ref{fig:cylbs}, where we have interpreted the periodic square grid pattern that runs across the entire $(k_{1\parallel}, k_{2\parallel})$ plane as arising from the periodic pattern of missing frequency channels. 

We now consider the bottom panels of Figure~\ref{fig:wedge_sq} to compare the amplitudes of $\mid B(k_{1\perp}, k_{1\parallel}) \mid$ at different $k_{1\parallel}$. In both the left and middle panels,  the range $k_{1\parallel} \approx 0$ is in the A0 region, and we have a peak value of $\sim 10^{24}~\rm{mK^3 ~Mpc^6}$  for both panels. The range $k_{1\parallel} \ge  0.5~{\rm Mpc}^{-1}$ is in the A3 region for the left panel, while it is in the A2 region in the middle panel. Leaving aside the spikes, we see that $\mid B(k_{1\perp}, k_{1\parallel}) \mid$ has values  $\sim 10^{16}-10^{17} \, \rm{mK^3~Mpc^6}$ in the left panel, whereas it has values 
$\sim 10^{19} \, \rm{mK^3~Mpc^6}$ in the right panel. The amplitude of the spikes is also smaller in the left panel $(\sim 10^{19}~\rm{mK^3~Mpc^6})$ compared to the right panel $(\sim 10^{21}~\rm{mK^3~Mpc^6})$.  

Figure~\ref{fig:wedge_sq}  clearly demonstrates a foreground wedge for the cylindrical BS, very similar to the cylindrical PS. We see that the level of foreground contamination successively decreases from region A0 to A3. We may interpret the A3 region, which has the least level of foreground contamination, as the EoR window for the cylindrical BS. However, the present data shows considerable leakage into the EoR window, and the situation is particularly aggravated by the periodic pattern of missing frequency channels that produces a periodic pattern of spikes across the EoR window. 

\subsection{The Spherical BS}

In the final step of our analysis, we calculate the 3D spherical BS $\bar{B}(k_1,k_2,k_3)$ with $k_1 \ge k_2 \ge k_3$, and where $k_1=\sqrt{k_{1\perp}^2 + k_{1\parallel}^2}$, etc.  Here, $(k_1,k_2,k_3)$ specify the lengths of the three sides of a triangle $(\kk_1,\kk_2,\kk_3)$, which uniquely specify its shape and size.  Note that it is necessary to also specify the orientation of the triangle with respect to the LoS in order to specify  $B(k_{1\perp}, k_{2\perp}, k_{3\perp}, k_{1\parallel}, k_{2\parallel})$ the cylindrical BS. 

Here we adopt a convenient parametrization \citep{bharad2020} for the triangle 
$(k_1,k_2,k_3)$, which uses the largest side $k_1$ to quantify the size and the two dimensionless parameters 
\begin{equation}
\mu =- \frac{\kk_1 \cdot \kk_2}{k_1 k_2}, \hspace{0.5cm} \, t = \frac{k_2}{k_1} \,.
\label{eq:shape}
\end{equation}
to quantify the shape. The values of the parameters $(\mu,t)$ are confined to the range 

\begin{equation}
    0.5 \leq \mu, t \leq 1 \hspace{0.5cm}  {\rm and} \hspace{0.5cm}  2\mu t \geq 1 \,.
    \label{eq:prange}
\end{equation}

 Furthermore, $k_3=k_1 \, \sqrt{1 + t^2 - 2 \mu t}$. The orientation of the triangle  with respect to the  LoS is specified by a unit vector $\pp$, for which 
\begin{equation}
    k_{1 \, \parallel}  = k_1 \, p_z \,,  \hspace{0.5 cm} \, 
k_{2 \, \parallel}  = k_1 t \, (-\mu \, p_z + \sqrt{1-\mu^2} \, p_x)  \,.
\label{eq:orient}
\end{equation}
In this parameterization, the cylindrical BS can be expressed as, 
\begin{equation}
    B(k_1,\mu,t,\pp) \equiv B(k_{1 \perp},k_{2 \perp},k_{3 \perp},k_{1\parallel},k_{2\parallel})  
    \label{eq:cvs}
\end{equation}
which is a function of five independent parameters. 

Redshift space distortion  (RSD) of the 21-cm signal introduces an anisotropy with respect to the LoS \citep{Bharadwaj_2004}, which causes   
the BS to depend on the orientation of the triangle. It is possible to quantify this by decomposing the BS in terms of spherical harmonic $Y_l^m(\pp)$  \citep{sco1999,bharad2020}, and we have the BS multipoles   
\begin{equation}
    \bar{B}_l^m(k_1,\mu,t )=\sqrt{\dfrac{2l+1}{4\pi}}{\int[Y_l^m(\hat{\textbf{p}})]^*~B(k_1,\mu,t,\pp )~d\Omega_{\hat{\textbf{p}}}} \,.
\label{eq:bs_lm}
\end{equation}
Theoretical predictions for the higher multipoles have been analyzed earlier 
\citep{bharad2020,Mazumdar_2020}. In a recent work, \citet{gill_2024} have implemented a brisk estimator for the angular multipoles of the BS. The present work follows \citet{gill_2024} for a discrete implementation of Eq.~(\ref{eq:bs_lm}), however the analysis here is restricted to $\bar{B}_0^0(k_1, \mu, t)$ the monopole, which we denote using 
 \begin{equation}
  \bar{B}(k_1, \mu, t) \equiv  \bar{B}_0^0(k_1, \mu, t) \,,
  \label{eq:sphbs}
 \end{equation} 
  and refer to as the spherical BS in the rest of this paper. For each estimate of the spherical BS, we also compute $N_{\rm tri}(k_1, \mu, t)$, the number of $(\kk_1,\kk_2,\kk_3)$ triangles that contribute to the estimate.

The estimated values of $\bar{B}(k_1, \mu, t)$ are irregularly distributed and do not uniformly sample $(k_1,\mu, t)$.
We perform an additional binning in the $(k_1, \mu, t)$ space to obtain a more uniform coverage and also enhance the signal-to-noise ratio. The $k_1$ range $[0.002~{\rm Mpc}^{-1}, 4.624 ~{\rm Mpc}^{-1}]$, is divided into 20 equal logarithmically spaced bins, while the $\mu$ and $t$ ranges are each divided into $5$ linearly spaced bins. The bin averaged values of $k_1$ span the range $[0.005~{\rm Mpc}^{-1} , 3.88 ~{\rm Mpc}^{-1}]$.  Within each bin, we compute a weighted average of the estimated spherical BS using the number of triangles as weights. 
We then have uniformly spaced estimates of $\bar{B}(k_1, \mu, t)$ and $N_{\rm tri}(k_1, \mu, t)$. Instead of the spherical BS, we consider the mean cube brightness temperature fluctuations
\begin{equation}
    \Delta^3(k_1,\mu,t)={(2\pi^2)^{-2}} \, {k_1^6 \, \bar{B}(k_1,\mu,t)} \,,
\label{eq:mcb}
\end{equation}
for which the results are shown in the next section. We have used noise-only simulations to predict the expected statistical fluctuations in the estimated values of $ \Delta^3(k_1,\mu,t)$ shown here.

\section{Results}
\label{sec:results}

 \begin{figure*}
\centering
\includegraphics[width=1\textwidth]{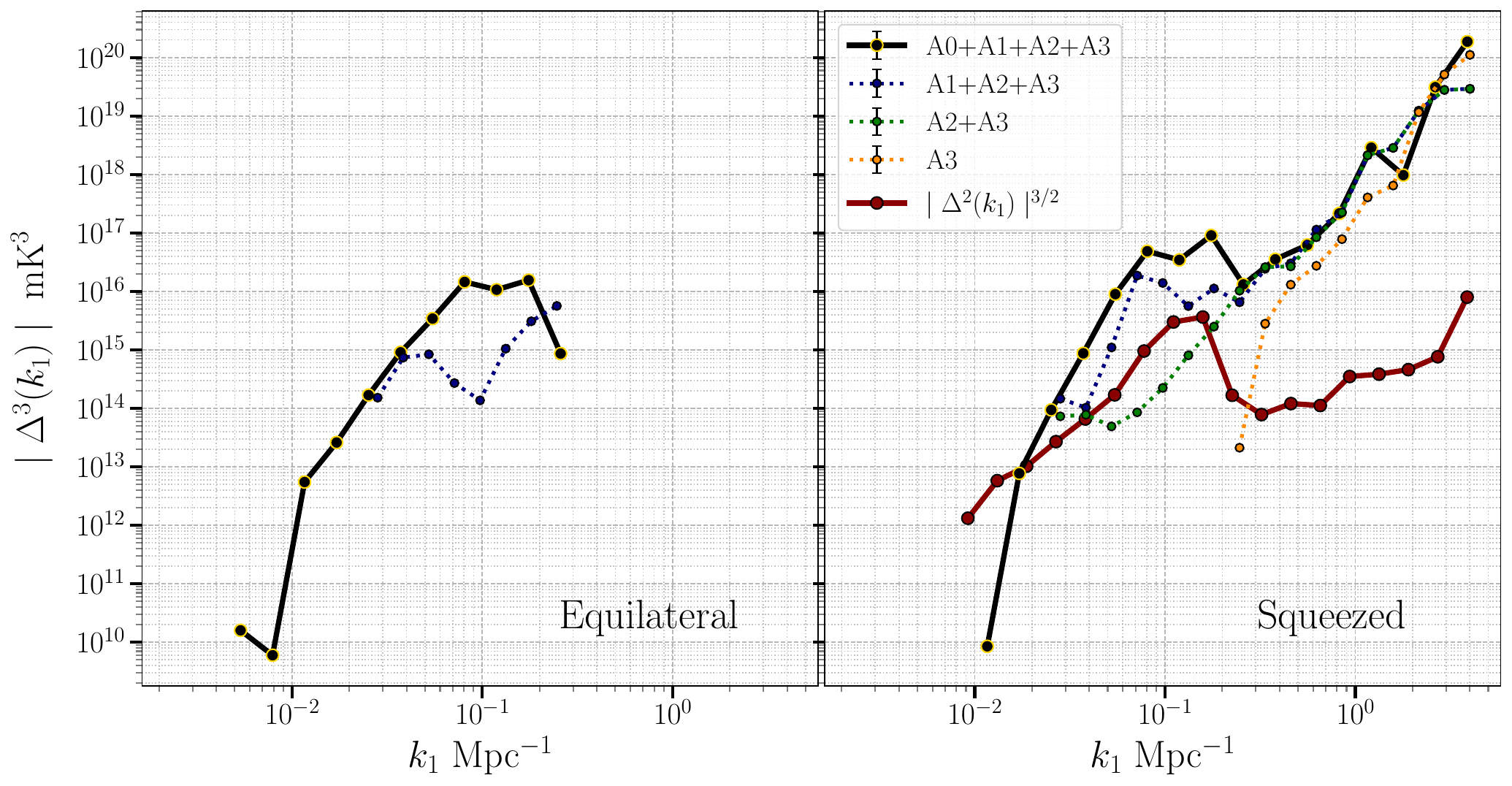}
\caption{Mean cube brightness temperature fluctuations, $\Delta^3(k_1)$, for equilateral triangles $(k_1 \approx k_2 \approx k_3)$ in the left panel and squeezed triangles $(k_1 \approx k_2,  k_3 \rightarrow 0)$ in the right panel. The solid black lines show results with no 
foreground avoidance (A0+A1+A2+A3). The dotted colored lines show results for various foreground avoidance scenarios, as indicated in the figure legend. The solid red line shows $|\Delta^2(k_1)|^{3/2}$, where $\Delta^2(k_1)$ is the mean square brightness temperature fluctuation from \citet{Elahi_missing} who have analyzed the same data.}
\label{fig:eq_sq}
\end{figure*}

Figure~\ref{fig:eq_sq} shows $|\Delta^3(k_1)|$, which corresponds to $|\Delta^3(k_1,\mu,t)|$ as a function of $k_1$ for fixed $(\mu,t)$. Here we show results for two fixed values $(\mu,t) \approx (0.5,1)$ and $(1,1)$, which respectively correspond to equilateral triangles (left panel) and squeezed triangles (right panel).    The solid black curve (A0+A1+A2+A3) shows $|\Delta^3(k_1)|$ estimated using all available values of  $B(k_{1 \perp},k_{2 \perp},k_{3 \perp},k_{1\parallel},k_{2\parallel})$, including those in the A0 region, which we know for sure to have a high level of foreground contamination.  We see that the estimates span the range $k_1 \in [0.005,\ 0.258]\,{\rm Mpc}^{-1}$ and $k_1 \in [0.012,\ 3.88]\,{\rm Mpc}^{-1}$ for equilateral and squeezed triangles, respectively. Note that we don't have squeezed triangles $(k_3 \ll k_1 \approx k_2)$ for a few of the smallest $k_1$. Furthermore, the large $k$ are mainly dominated by $k_{\parallel}$  as against $k_{\perp}$, and the condition 
 $k_{3\parallel} = -(k_{1\parallel} + k_{2\parallel})$ precludes equilateral triangles for large $k_1$.  We find that $|\Delta^3(k_1)|$ shows very similar behavior for equilateral and squeezed triangles. In both cases,
  $|\Delta^3(k_1)| \approx  10^{10}~{\rm mK}^3$ at the smallest $k_1$, 
  the amplitude increases monotonically with $k_1$ until $k_1 \approx 0.2~{\rm Mpc}^{-1}$ where  $|\Delta^3(k_1)| \approx  10^{16}~{\rm mK}^3$  and $10^{17}~{\rm mK}^3$ for equilateral and squeezed triangles, respectively. Beyond this, 
  in both cases, $|\Delta^3(k_1)|$ drops abruptly by one order of magnitude at  
  $k_1 \approx 0.3~{\rm Mpc}^{-1}$. The equilateral triangles do not extend to larger $k_1$, for squeezed triangles,  $|\Delta^3(k_1)|$ continues to increase and reaches $ \approx  10^{20}~{\rm mK}^3$ at $k_1 \approx 4 ~{\rm Mpc}^{-1}$. Note the abrupt change in the behavior of  $|\Delta^3(k_1)|$ at $k_1 \approx 0.3~{\rm Mpc}^{-1}$, not only does the value fall by an order of magnitude, but the slope also changes at larger $k_1$. 

The solid red curve in the right panel of  Figure~\ref{fig:eq_sq} shows  
$|\Delta^2(k_1)|^{3/2}$, where the mean square brightness temperature fluctuation
$\Delta^2(k_1)={(2\pi^2)}^{-1} \,  {k_1^3 \, P(k_1)}$ is calculated using $P(k_1)$ 
the spherical PS estimated using all the $(k_{1\perp},k_{1\parallel})$ modes shown in Figure~\ref{fig:wedge_sq}. We see that at small $k_1$, the values of 
$|\Delta^2(k_1)|^{3/2}$ are comparable to those of $|\Delta^3(k_1)|$. In addition,  $|\Delta^2(k_1)|^{3/2}$ and $|\Delta^3(k_1)|$ both show very similar behavior in which the values increase with $k_1$ until $k_1 \approx 0.2~{\rm Mpc}^{-1}$, then drop abruptly at $k_1 \approx 0.3~{\rm Mpc}^{-1}$ and then increase again with a different slope.  The large $k_1$ modes are predominantly outside the foreground wedge, and in all cases, we may associate the transition 
at $k_1 \approx 0.2 - 0.3 \, ~{\rm Mpc}^{-1}$ with a crossing of the wedge boundary (Figure~\ref{fig:wedge_sq}).  For both the PS and the BS, the results are strongly foreground dominated at  $k_1 \le 0.2  \, ~{\rm Mpc}^{-1}$, while we have a significant number of modes (triangles) in the EoR window for larger $k_1$.

We next attempt to mitigate foreground contamination in the estimated spherical BS by adopting the foreground avoidance technique. We have seen that A0 is most severely affected by foregrounds. In both panels, the blue dotted line  (A1+A2+A3) shows $|\Delta^3|$ when A0 is excluded. This eliminates the estimates at small $k_1$, and now we only have estimates of  $|\Delta^3|$ for  $k_1 > 0.02~{\rm Mpc}^{-1}$.  We find some reduction in the values of  $|\Delta^3|$ in the range $k_1 < 0.3~{\rm Mpc}^{-1}$, but not beyond. In the subsequent cases, we only have estimates of $|\Delta^3|$ for squeezed triangles. 
 For the green dotted curve (A2+A3),   there is some further reduction at $k_1 < 0.3~{\rm Mpc}^{-1}$, but not beyond. The yellow dotted curve (A3) shows the most conservative situation, which only considers triangles whose three sides are all outside the foreground wedge. At $k_1 \approx 0.3~{\rm Mpc}^{-1}$, which is the smallest $k_1$ where we have an estimate, we find a substantial reduction in the value of  $|\Delta^3|$ from $1.31\times 10^{16}~{\rm mK}^3$ (A0+A1+A2+A3) to $2.12\times 10^{13}~{\rm mK}^3$  (A3). In this case, foreground avoidance reduces the value of the spherical BS  by a factor of $\approx 620$, that is, a reduction in foreground contamination by nearly three orders of magnitude. 
This large reduction indicates that at this $k_1$,  prior to foreground avoidance, $|\Delta^3|$   has a substantial contribution from A0, A1, and A2 triangles that contain a large level of foreground contamination.  We avoid the contribution from these triangles when we consider only A3. 
Considering larger $k_1$ ($>0.3~{\rm Mpc}^{-1}$), here also we find a reduction in the values of $|\Delta^3|$, however, the effect is relatively small. This indicates that the triangles here are mostly in the A3 region, both before and after foreground avoidance. 

The $1\sigma$ error bars corresponding to system noise have been shown for all the results in Figure~\ref{fig:eq_sq}. However, these are too small to be visible here. Based on this, we conclude that all the BS (or $|\Delta^3|$) values shown here are foreground-dominated. Although foreground avoidance leads to a substantial improvement in at least one case shown here, it fails to completely eliminate foreground contamination from the final estimated value. This indicates substantial leakage into the EoR window, as visible in 
Figure~\ref{fig:wedge_sq}. This issue is particularly aggravated by the periodic pattern of missing frequency channels.

 \begin{figure*}
\centering
\includegraphics[width=1\textwidth]{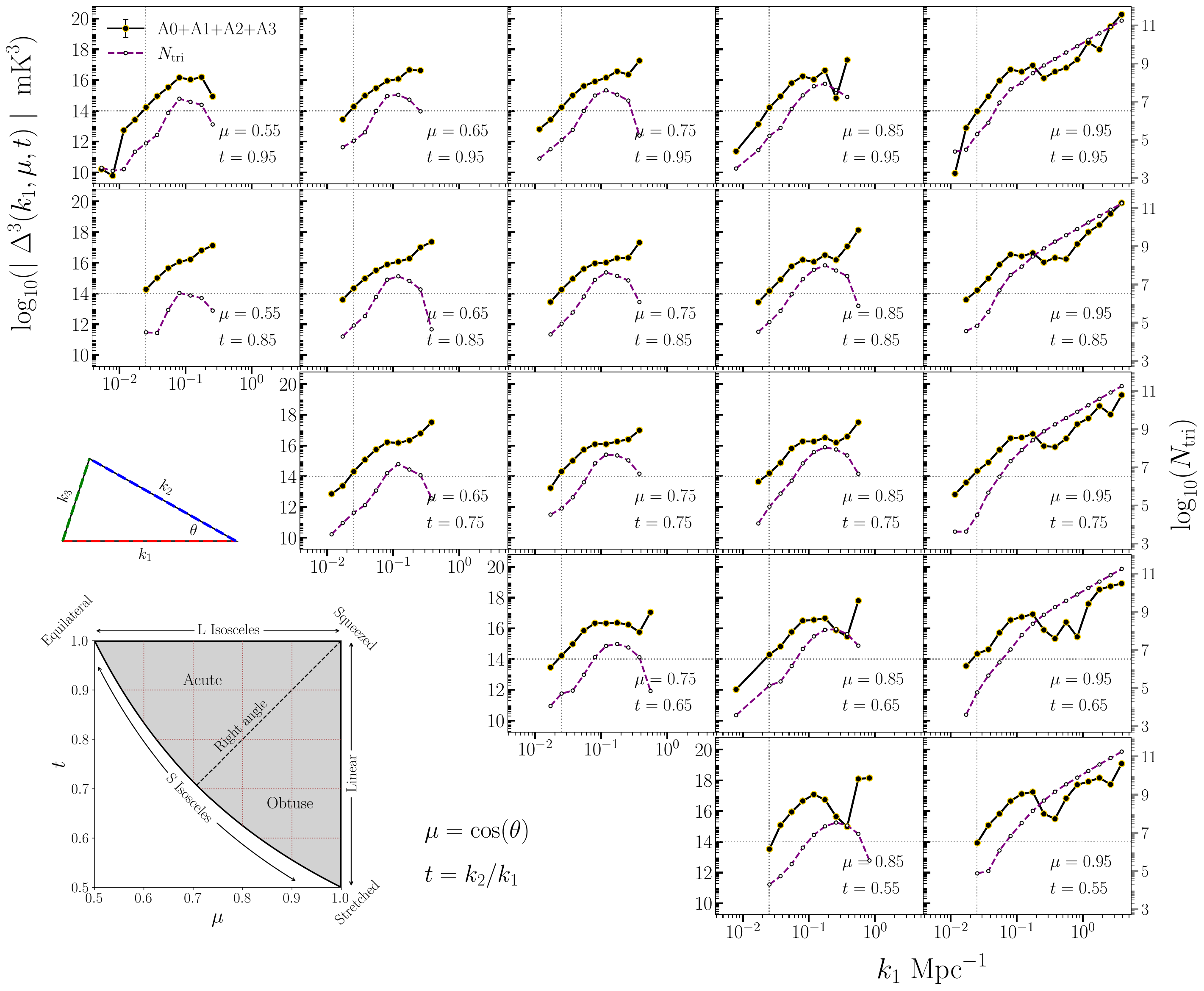}
\caption{ $\mid \Delta^3(k_1,\mu,t) \mid $  for triangles of all possible sizes $(k_1)$
and shapes $(\mu,t)$ available from this data. Each panel, which shows $\mid \Delta^3(k_1) \mid $ as a function of $k_1$, corresponds to a different shape $(\mu,t)$, as indicated by the inset figure in the lower left corner, and detailed in \citet{Bharadwaj_2019}. 
 The top-left and top-right panels, respectively,  correspond to the equilateral $(\mu=0.55, t=0.95)$ and squeezed $(\mu=0.95, t=0.95)$ triangles for which the results have been shown in 
Figure~\ref{fig:eq_sq}. The results, shown in black solid lines, do not incorporate any foreground avoidance (A0+A1+A2+A3).  The black dashed lines,  $k_1 = 0.025~\mathrm{Mpc}^{-1}$ (vertical) and  $|\Delta^3| = 10^{14}~\mathrm{mK}^3$ (horizontal), are shown for reference.
 The purple dashed lines show $N_{\rm tri}(k_1, \mu, t)$, the number of $(\kk_1,\kk_2,\kk_3) $ triangles in each $(k_1, \mu, t)$ bin. }
\label{fig:3dbs}
\end{figure*}

Figure~\ref{fig:3dbs} shows $|\Delta^3(k_1,\mu,t)|$ as a function of $k_1$, where each panel corresponds to a different value of $(\mu,t)$. As mentioned earlier, the allowed range of $(\mu,t)$  (Eq.~\ref{eq:prange}) has been divided into $5 \times 5$ bins, not all of which are within the allowed region shown in the reference plot at the bottom left corner of the figure.  Each panel corresponds to a different triangle shape,  as indicated in the reference plot and discussed in detail in  \citet{bharad2020}. Note, the isosceles triangles have been classified as L and S depending on whether their two large sides or small sides are equal. The equilateral and squeezed triangles, which have already been discussed in detail, are shown respectively in the top left and right panels.  
The bottom right panel shows the results for stretched triangles where $k_1/2 \approx k_2 \approx k_3$.

The black solid lines in the different panels of Figure~\ref{fig:3dbs} 
show $|\Delta^3(k_1,\mu,t)|$ obtained by combining all available triangles 
(A0+A1+A2+A3).  The $k_1$ range across which we have estimates of $|\Delta^3|$
varies with $(\mu,t)$. We see that the results shown in the different panels are broadly very similar to those shown in Figure~\ref{fig:eq_sq}, which has already been discussed in detail. At small $k_1$,  $|\Delta^3(k_1)|$  increases with $k_1$, then exhibits a dip around $k_1 \approx 0.2$–$0.3 ~ {\rm Mpc}^{-1}$ and then increases again with a different slope at larger $k_1$. Each panel shows two reference lines, one vertical at $k_1 = 0.025~\mathrm{Mpc}^{-1}$ and the other horizontal at $|\Delta^3| = 10^{14}~\mathrm{mK}^3$. These lines illustrate that at small $k_1$,  
 $|\Delta^3|$  has roughly the same value in all the panels.  In most cases, $|\Delta^3|$ ranges from $10^{13}$ to $10^{17}~\mathrm{mK}^3$ at small  $k_1$.  Note that the available $k_1$ range depends on $(\mu,t)$ (the triangle shape), with the squeezed triangle spanning the widest $k_1$ range, followed by the other linear (degenerate)  triangle along $\mu \approx 1$. The value of $|\Delta^3|$ reaches $ \approx  10^{20}~{\rm mK}^3$ at $k_1 \approx 4 ~{\rm Mpc}^{-1}$, which is the largest value of $k_1$  here.  

 The purple dashed line in each panel shows $N_{\rm tri}(k_1,\mu,t)$, which is the number of $(\kk_1,\kk_2,\kk_3)$  triangles contained in the corresponding $(k_1,\mu,t)$ 
bin. We see that $N_{\rm tri}$ spans a wide range of values from $\sim 10^{3}$ to $\sim 10^{11}$, as indicated on the right side of the figure. In all cases,    
$N_{\rm tri}$ is small at low $k_1$ and initially increases with $k_1$. For most triangle shapes,  excluding the linear ones $(\mu \approx 1)$,  
$N_{\rm tri}$ peaks at $\sim 10^8$ around $k_1 \approx 0.1$–$0.3~{\rm Mpc}^{-1}$ and subsequently decreases at larger $k_1$. For $\mu \approx 1$, the value of $N_{\rm tri}$ increases monotonically to $\sim 10^{11}$ at the largest $k_1$.

 \begin{table*}[ht]
  \centering
  \caption{The measured $\Delta^3(k_1,\mu,t)$, corresponding $1\sigma$ systematical uncertainties, $2\sigma$ upper limit $\Delta^3_{\rm UL}$, and number of triangles $N_{\rm tri}$. The values are shown only for two triangle configurations—equilateral and squeezed—where the tightest upper limits $\Delta^3_{\rm UL}$ are obtained. The full set of estimates for all triangle types is provided in Table~\ref{tab:bs_estimates} in the Appendix.}
  \label{tab:bs_eqsq}
  \begin{tabular}{l c c c c c}
    \toprule
    Shape 
      & \multicolumn{1}{c}{Size} 
      & \multicolumn{3}{c}{Bispectrum} 
      & Triangles \\
    \cmidrule(lr){2-2}\cmidrule(lr){3-5}\cmidrule(lr){6-6}
      & $k_{1}\,\mathrm{[Mpc^{-1}]}$ 
      & $\Delta^{3}~[\rm mK^3]$ 
      & $1\sigma~[\rm mK^3]$ 
      & $\Delta^{3}_{\rm UL}~[\rm mK^3]$ 
      & $N_{\rm tri}$ \\
    \midrule
    Equilateral $(\mu=0.5,~t=0.95)$ 
      & $0.008$ 
      & $(1.81\times10^{3})^{3}$ 
      & $(3.48\times10^{1})^{3}$ 
      & $(1.81\times10^{3})^{3}$ 
      & $2.38\times10^{3}$ \\
    Squeezed  $~(\mu=0.95,~t=0.95)$    
      & $0.012$ 
      & $(2.04\times10^{3})^{3}$ 
      & $(4.74\times10^{1})^{3}$ 
      & $(2.04\times10^{3})^{3}$ 
      & $2.44\times10^{4}$ \\
    \bottomrule
  \end{tabular}
\end{table*}

The $1 \sigma$ error bars shown in all the panels of Figure~\ref{fig:3dbs}  are too small to be visible, and we interpret the results shown here as foreground-dominated. We use these foreground-dominated estimates to place upper limits on $\Delta^3$ for the EoR 21-cm signal. We find the most stringent upper limits at the smallest values of $k_1$
for equilateral and squeezed triangles, which are tabulated in 
 Table~\ref{tab:bs_eqsq}.  The $2 \sigma$ upper limit $\Delta^3_{\rm UL}$ has values  
 $(1.81 \times 10^3)^3~\mathrm{mK}^3$ at $k_1 = 0.008~\mathrm{Mpc}^{-1}$  and 
 $(2.04 \times 10^3)^3~\mathrm{mK}^3$ at $k_1 = 0.012~\mathrm{Mpc}^{-1}$ for equilateral and squeezed triangles respectively.   The complete set of estimates and upper limits for all $|\Delta^3(k_1,\mu,t)|$ shown in the different panels of Figure~\ref{fig:3dbs} is provided in Table~\ref{tab:bs_estimates} in the Appendix.

 The $z \approx 8$ EoR 21-cm signal is predicted to be $\Delta^3 \approx 10^3$ to $30^3~\mathrm{mK}^3$ within the range $k_1 \approx 0.29$ to $2.48~\mathrm{Mpc}^{-1}$ \citep{gill_eormulti}, which is approximately $13-15$ orders of magnitude smaller than the upper limits obtained here in the same $k_1$ range.   As discussed in Figure~\ref{fig:eq_sq}, for a few $k_1$ bins, it is possible to reduce the upper limits on $\Delta^3$  by $2-3$ orders of magnitude if we adopt foreground avoidance. In the present work, 
 we have not considered foreground removal for the different $(\mu,t)$ bins shown in Figure~\ref{fig:3dbs}. In the future, we plan to consider the issue of foreground mitigation in more detail.

\section{Summary and Conclusions}
\label{sec:sum}

The EoR 21-cm signal is predicted to be highly non-Gaussian due to the presence of ionized bubbles  \citep{BP2005}. The power spectrum (PS), which is commonly employed to characterize the statistics of the EoR 21-cm signal, is insensitive to non-Gaussianity. The bispectrum (BS) is the lowest-order statistic that is sensitive to non-Gaussianity. Here, we have used a recently developed visibility-based estimator (\citetalias{Gill_2025_mabs}) to compute the $z=8.2$ EoR 21-cm BS from a single pointing of the MWA drift-scan observations \citep{Patwa_2021}. 

In the first step of our analysis, we estimate the multi-frequency angular bispectrum (MABS) $B_A(\ell_1, \ell_2, \ell_3, \Delta\nu_1, \Delta\nu_2)$ from gridded visibility data. The MABS, shown in Figures \ref{fig:mabs} and \ref{fig:mabs_slice}, is foreground-dominated, and it has a  peak value  $\sim 10^6~{\rm mK}^3$ at $\Delta\nu_1= \Delta\nu_2=0$. The value of the MABS falls at larger $(\Delta\nu_1, \Delta\nu_2)$,  and it drops to  approximately $1/6$ the peak value at the largest frequency separation 
$ \pm 15.36 \, {\rm MHz}$.  This behavior of the foreground signal is quite different from that expected for a cosmological 21-cm signal  (\citetalias{Gill_2025_mabs}) where the MABS is predicted to de-correlate rapidly with increasing frequency separation $(\Delta\nu_1, \Delta\nu_2)$, and have a value close to zero at frequency separations larger than $ \pm 2 \, {\rm MHz}$. Although the MWA data has a periodic pattern of missing frequency channels, we do not find any missing frequency separation $(\Delta\nu_1, \Delta\nu_2)$.  However, as noted in \citetalias{Gill_2025_mabs}, this introduces a periodic modulation in the sampling of the  $(\Delta\nu_1, \Delta\nu_2)$ values.

We have calculated the 3D cylindrical BS $B(k_{1\perp}, k_{2\perp}, k_{3\perp}, k_{1\parallel}, k_{2\parallel})$  by performing a 2D Fourier transform 
of the MABS with respect to $(\Delta\nu_1, \Delta\nu_2)$. The values of $B(k_{1\perp}, k_{2\perp}, k_{3\perp}, k_{1\parallel}, k_{2\parallel})$, shown in Figure \ref{fig:cylbs}, peak around  $k_{1\parallel}= k_{1\parallel}=0$, and  
decline rapidly with increasing $(k_{1\parallel}, k_{2\parallel})$. 
In addition,  we observe a square grid pattern that runs across the entire $(k_{1\parallel}, k_{2\parallel})$ plane. This 2D oscillatory pattern has a period of $\delta k_\parallel=0.29~{\rm Mpc}^{-1}$, which exactly matches the period of the pattern of missing frequency channels in the MWA data.
Furthermore, we notice three bands that run across the $(k_{1\parallel}, k_{2\parallel})$ plane, which we identify as the foreground wedge corresponding to $(k_{1\perp},k_{1 \parallel})$, $(k_{2\perp},k_{2 \parallel})$ and $(k_{3\perp},k_{3 \parallel})$ respectively. Based on this, we partition the entire $(k_{1\parallel}, k_{2\parallel})$ plane into four distinct regions, namely A0, A1, A2 and A3. In A3, all the three modes $(k_{1\perp},k_{1 \parallel})$, $(k_{2\perp},k_{2 \parallel})$ and $(k_{3\perp},k_{3 \parallel})$ avoid the respective foreground wedge, whereas only two of the modes avoid the foreground wedge for A2, and so forth. We expect A3  to have the lowest level of foreground contamination in the cylindrical BS $B(k_{1\perp}, k_{2\perp}, k_{3\perp}, k_{1\parallel}, k_{2\parallel})$, and this is expected to increase as we go from A3 to A0.  

Figure \ref{fig:wedge_sq}, which shows $B(k_{1\perp},k_{1\parallel})$ the cylindrical BS for two specific triangle configurations, clearly illustrates the foreground wedge. This figure also shows  $P(k_{1\perp},k_{1\parallel})$, the cylindrical PS from \citet{Elahi_missing}, for exactly the same MWA data. 
The familiar foreground wedge is clearly visible in the $(k_{1\perp},k_{1\parallel})$ plane  for both 
$B(k_{1\perp},k_{1\parallel})$  and  $P(k_{1\perp},k_{1\parallel})$. 
 The BS $B(k_{1\perp},k_{1\parallel})$ clearly shows the expected differences in the levels of foreground contamination in the regions A0, A1, A2, and A3.  We identify A3, which has the lowest level of foreground contamination, as the EoR window for the BS. Both the PS and the BS exhibit a periodic pattern of spikes due to the periodic pattern of missing frequency channels. This extends across the entire $(k_{1\perp},k_{1\parallel})$ plane, causing severe foreground leakage into the EoR window.  Although we have illustrated the foreground wedge for only two specific triangle shapes here, we have verified that this feature is present across all possible triangle shapes.

The present analysis includes all possible triangles $(\kk_1,\kk_2,\kk_3)$ that arise from the given data, which is a total number of $N_{\rm tri}=1.66\times10^{12}$  triangles. Of these, approximately $83.6\%$   are in the EoR window (A3 region, Figure \ref{tab:ntri_by_region_horizontal}). This bodes well for using foreground avoidance to mitigate foreground contamination for the 21-cm BS. However, we do not get any estimates at small $k_1$ if we restrict the analysis to the EoR window. 

We finally use the measured $B(k_{1\perp}, k_{2\perp}, k_{3\perp}, k_{1\parallel}, k_{2\parallel})$ to evaluate the 3D spherical BS $\bar{B}(k_1,\mu,t)$, where $k_1$ and $(\mu,t)$, respectively, quantify the size and shape of the triangle $(k_1,k_2,k_3)$.  Here we consider the mean cube brightness temperature fluctuations $\Delta^3(k_1,\mu,t)={(2\pi^2)^{-2}} \, {k_1^6 \, \bar{B}(k_1,\mu,t)}$, and we present results for $\Delta^3(k_1)$ as a function of $k_1$, considering triangle of all possible shapes $(\mu,t)$. 
Figure \ref{fig:eq_sq}, which considers a variety of scenarios for foreground avoidance,  shows $\Delta^3(k_1)$ for two particular triangle shapes, namely equilateral and squeezed. Figure~\ref{fig:3dbs} shows $\Delta^3(k_1)$ for triangles of all possible shapes, without incorporating foreground avoidance.    We find that the available $k_1$ range varies with the triangle shape and the foreground avoidance scenario. We first consider the results without foreground avoidance (A0+A1+A2+A3). We have the broadest $k_1$ range $[0.012~{\rm Mpc}^{-1}, 3.88~{\rm Mpc}^{-1}]$ for squeezed triangles, although the lowest $k_1$ value  $(0.005~{\rm Mpc}^{-1})$   is achieved for equilateral triangles, for which the large $k_1$  $( >0.3~{\rm Mpc}^{-1})$ are missing. We have the lowest value of $|\Delta^3(k_1)| \approx  10^{10}~{\rm mK}^3$ at the smallest $k_1$ for equilateral and squeezed triangles (Table~\ref{tab:bs_eqsq}).  The lowest values of $|\Delta^3(k_1)|$ 
are in the range  $10^{12}$–$10^{14}\,{\rm mK}^3$ for other shapes. 
We find the largest value of $|\Delta^3(k_1)| \approx 10^{20} \,{\rm mK}^3$ at   $3.88 ~{\rm Mpc}^{-1}$  for squeezed triangles. In this study, we have carried out simulations that only incorporate the expected system noise contribution to the visibilities and analyzed these in exactly the same way as the actual data. 
These noise-only simulations were used to calculate $\sigma$ the r.m.s.  statistical fluctuations expected in the estimated $\Delta^3$.  
The values of $\Delta^3$, $\sigma$ and the $2 \sigma$ upper limits  $\Delta^3_{\rm UL} $  for all $(k_1,\mu,t)$ are tabulated in Table~\ref{tab:bs_estimates} in the Appendix.
The estimated values of $\Delta^3$  are several orders of magnitude larger than those of $\sigma $, which leads us to believe that the present estimates are foreground dominated. We proceed to use these estimates to constrain the EoR 21-cm BS. The most stringent $2\sigma$ upper limits that we obtain are $\Delta^3_{\rm UL} = (1.81\times 10^3)^3~\mathrm{mK}^3$ at $k_1 = 0.008~\mathrm{Mpc}^{-1}$, and $\Delta^3_{\rm UL} = (2.04\times 10^3)^3~\mathrm{mK}^3$ at $k_1 = 0.012~\mathrm{Mpc}^{-1}$ for equilateral and squeezed triangles respectively. Simulations predict the $z \approx 8$ EoR 21-cm signal to be  $\mid \Delta^3 \mid  \approx 10^3$ to $30^3~\mathrm{mK}^3$ within the range $k_1 \approx 0.29$ to $2.48~\mathrm{Mpc}^{-1}$ \citep{gill_eormulti}. Furthermore, $\Delta^3$ is predicted to be negative 
for small $k_1$ ($\lesssim  0.56~\mathrm{Mpc}^{-1}$), which is an important signature of the distinct ionized bubbles that are present in the neutral  \Hi~background. The predicted EoR 21-cm BS is roughly 13 to 15 orders of magnitude smaller than the current upper limits obtained in the same $k_1$ range.

We have explored various foreground avoidance scenarios, for which the results for equilateral and squeezed triangles are shown in Figure \ref{fig:eq_sq}. With foreground avoidance, we do not have estimates of $\Delta^3$ at small $k_1$. As a consequence, there is very little scope for using this technique for equilateral triangles.  For squeezed triangles, considering only the EoR window, we have a substantial reduction (a factor of $\approx 620$ in the foreground contamination at the smallest $k_1$  $(\approx 0.3~{\rm Mpc}^{-1})$. However, we only have a small reduction at larger $k_1$, most probably due to the strong leakage into the EoR window due to the periodic pattern of missing frequency channels.

We plan to present a more comprehensive investigation of foreground mitigation strategies in future work. Furthermore, it may be advantageous first to remove the dominant foreground components by either modeling the sky image or using some frequency-based technique like  Smooth Component Filtering \citep{Elahi_missing} prior to estimating the BS. We also plan to analyze all the pointing in the drift scan observations, once we achieve some control over the foreground contamination.

\section*{Acknowledgments}
We acknowledge the computing facilities in the Department of Physics, IIT Kharagpur.

\section*{Data Availability}

The data sets were derived from sources in the public domain
(the MWA Data Archive: project ID G0031) available at \href{https://asvo.mwatelescope.org/}{MWA ASVO Portal}.

\clearpage 
\appendix
\label{ap:table}

Table~\ref{tab:bs_estimates} tabulates all the values of $\Delta^3(k_1,\mu,t)$ shown in the different  panels of  Figure~\ref{fig:3dbs}. The corresponding values of $\sigma$, $\Delta^3_{\rm UL}$ and $N_{\rm tri}$ are shown alongside the values of $\Delta^3$. 

\begin{longtable}{l ll cccc}
\caption{Measurements of $\Delta^3$ and upper limit $\Delta^3_{\rm UL}$ for all possible triangle configurations $(k_1,\mu,t)$.}\label{tab:bs_estimates}\\
\toprule
\multicolumn{1}{c}{\textbf{Size}} & \multicolumn{2}{c}{\textbf{Shape}} & \multicolumn{3}{c}{\textbf{Bispectrum}} & \multicolumn{1}{c}{\textbf{Triangles}} \\
\cmidrule(l){1-1} \cmidrule(l){2-3} \cmidrule(l){4-6} \cmidrule(l){7-7}
$k_1$\,[Mpc$^{-1}$] & $\mu$ & $t$ & $\Delta^3$\,[mK$^{3}$] & $1\sigma$\,[mK$^{3}$] & $\Delta^3_{\rm UL}$\,[mK$^{3}$] & $N_{\rm tri}$ \\
\midrule
\endfirsthead
\toprule
\multicolumn{1}{c}{\textbf{Size}} & \multicolumn{2}{c}{\textbf{Shape}} & \multicolumn{3}{c}{\textbf{Bispectrum}} & \multicolumn{1}{c}{\textbf{Triangles}} \\
\cmidrule(l){1-1} \cmidrule(l){2-3} \cmidrule(l){4-6} \cmidrule(l){7-7}
$k_1$\,[Mpc$^{-1}$] & $\mu$ & $t$ & $\Delta^3$\,[mK$^{3}$] & $1\sigma$\,[mK$^{3}$] & $\Delta^3_{\rm UL}$\,[mK$^{3}$] & $N_{\rm tri}$ \\
\midrule
\endhead
\midrule
\multicolumn{7}{r}{\textit{continued on next page}} \\
\endfoot
\bottomrule
\endlastfoot
0.005 & 0.55 & 0.95 & $-(2.52 \times 10^{3})^3$ & $(1.48 \times 10^{1})^3$ & $(2.52 \times 10^{3})^3$ &  $3.43\times 10^{3}$ \\
\midrule
0.008 & 0.85 & 0.65 &  $(1.02 \times 10^{4})^3$ & $(2.77 \times 10^{1})^3$ & $(1.02 \times 10^{4})^3$ &  $3.81\times 10^{3}$ \\
      & 0.55 & 0.95 &  $(1.81 \times 10^{3})^3$ & $(3.48 \times 10^{1})^3$ & $(1.81 \times 10^{3})^3$ &  $2.38\times 10^{3}$ \\
      & 0.85 & 0.95 &  $(6.16 \times 10^{3})^3$ & $(3.16 \times 10^{1})^3$ & $(6.16 \times 10^{3})^3$ &  $3.15\times 10^{3}$ \\
\midrule
0.012 & 0.65 & 0.75 &  $(1.93 \times 10^{4})^3$ & $(6.44 \times 10^{1})^3$ & $(1.93 \times 10^{4})^3$ &  $3.05\times 10^{3}$ \\
      & 0.95 & 0.75 &  $(1.88 \times 10^{4})^3$ & $(4.97 \times 10^{1})^3$ & $(1.88 \times 10^{4})^3$ &  $4.21\times 10^{3}$ \\
      & 0.55 & 0.95 &  $(1.76 \times 10^{4})^3$ & $(7.05 \times 10^{1})^3$ & $(1.76 \times 10^{4})^3$ &  $2.90\times 10^{3}$ \\
      & 0.75 & 0.95 &  $(1.85 \times 10^{4})^3$ & $(5.00 \times 10^{1})^3$ & $(1.85 \times 10^{4})^3$ &  $1.05\times 10^{4}$ \\
      & 0.95 & 0.95 &  $(2.04 \times 10^{3})^3$ & $(4.74 \times 10^{1})^3$ & $(2.04 \times 10^{3})^3$ &  $2.44\times 10^{4}$ \\
\midrule
0.017 & 0.75 & 0.65 &  $(3.07 \times 10^{4})^3$ & $(1.00 \times 10^{2})^3$ & $(3.07 \times 10^{4})^3$ &  $1.16\times 10^{4}$ \\
      & 0.95 & 0.65 &  $(3.30 \times 10^{4})^3$ & $(1.27 \times 10^{2})^3$ & $(3.30 \times 10^{4})^3$ &  $4.05\times 10^{3}$ \\
      & 0.65 & 0.75 &  $(2.88 \times 10^{4})^3$ & $(1.28 \times 10^{2})^3$ & $(2.88 \times 10^{4})^3$ &  $1.19\times 10^{4}$ \\
      & 0.75 & 0.75 &  $(2.58 \times 10^{4})^3$ & $(9.86 \times 10^{1})^3$ & $(2.58 \times 10^{4})^3$ &  $3.29\times 10^{4}$ \\
      & 0.85 & 0.75 &  $(3.54 \times 10^{4})^3$ & $(1.04 \times 10^{2})^3$ & $(3.54 \times 10^{4})^3$ &  $1.13\times 10^{4}$ \\
      & 0.95 & 0.75 &  $(3.41 \times 10^{4})^3$ & $(1.22 \times 10^{2})^3$ & $(3.41 \times 10^{4})^3$ &  $4.23\times 10^{3}$ \\
      & 0.65 & 0.85 &  $(3.41 \times 10^{4})^3$ & $(1.14 \times 10^{2})^3$ & $(3.41 \times 10^{4})^3$ &  $1.87\times 10^{4}$ \\
      & 0.75 & 0.85 &  $(3.05 \times 10^{4})^3$ & $(9.83 \times 10^{1})^3$ & $(3.05 \times 10^{4})^3$ &  $2.41\times 10^{4}$ \\
      & 0.85 & 0.85 &  $(3.03 \times 10^{4})^3$ & $(9.87 \times 10^{1})^3$ & $(3.03 \times 10^{4})^3$ &  $3.31\times 10^{4}$ \\
      & 0.95 & 0.85 &  $(3.41 \times 10^{4})^3$ & $(8.94 \times 10^{1})^3$ & $(3.41 \times 10^{4})^3$ &  $3.60\times 10^{4}$ \\
      & 0.55 & 0.95 &  $(2.96 \times 10^{4})^3$ & $(1.13 \times 10^{2})^3$ & $(2.96 \times 10^{4})^3$ &  $2.40\times 10^{4}$ \\
      & 0.65 & 0.95 &  $(3.03 \times 10^{4})^3$ & $(8.80 \times 10^{1})^3$ & $(3.03 \times 10^{4})^3$ &  $4.12\times 10^{4}$ \\
      & 0.75 & 0.95 &  $(2.97 \times 10^{4})^3$ & $(1.06 \times 10^{2})^3$ & $(2.97 \times 10^{4})^3$ &  $3.23\times 10^{4}$ \\
      & 0.85 & 0.95 &  $(2.38 \times 10^{4})^3$ & $(1.14 \times 10^{2})^3$ & $(2.38 \times 10^{4})^3$ &  $2.90\times 10^{4}$ \\
      & 0.95 & 0.95 &  $(1.97 \times 10^{4})^3$ & $(9.25 \times 10^{1})^3$ & $(1.97 \times 10^{4})^3$ &  $3.06\times 10^{4}$ \\
\midrule
0.025 & 0.85 & 0.55 &  $(3.23 \times 10^{4})^3$ & $(2.65 \times 10^{2})^3$ & $(3.23 \times 10^{4})^3$ &  $1.94\times 10^{4}$ \\
      & 0.95 & 0.55 &  $(4.40 \times 10^{4})^3$ & $(1.78 \times 10^{2})^3$ & $(4.40 \times 10^{4})^3$ &  $7.39\times 10^{4}$ \\
      & 0.75 & 0.65 &  $(5.48 \times 10^{4})^3$ & $(2.01 \times 10^{2})^3$ & $(5.48 \times 10^{4})^3$ &  $5.20\times 10^{4}$ \\
      & 0.85 & 0.65 &  $(5.80 \times 10^{4})^3$ & $(1.62 \times 10^{2})^3$ & $(5.80 \times 10^{4})^3$ &  $1.36\times 10^{5}$ \\
      & 0.95 & 0.65 &  $(6.03 \times 10^{4})^3$ & $(1.77 \times 10^{2})^3$ & $(6.03 \times 10^{4})^3$ &  $6.01\times 10^{4}$ \\
      & 0.65 & 0.75 &  $(5.88 \times 10^{4})^3$ & $(2.10 \times 10^{2})^3$ & $(5.88 \times 10^{4})^3$ &  $4.15\times 10^{4}$ \\
      & 0.75 & 0.75 &  $(5.89 \times 10^{4})^3$ & $(1.77 \times 10^{2})^3$ & $(5.89 \times 10^{4})^3$ &  $6.90\times 10^{4}$ \\
      & 0.85 & 0.75 &  $(5.43 \times 10^{4})^3$ & $(1.69 \times 10^{2})^3$ & $(5.43 \times 10^{4})^3$ &  $8.54\times 10^{4}$ \\
      & 0.95 & 0.75 &  $(6.05 \times 10^{4})^3$ & $(2.06 \times 10^{2})^3$ & $(6.05 \times 10^{4})^3$ &  $3.17\times 10^{4}$ \\
      & 0.55 & 0.85 &  $(5.72 \times 10^{4})^3$ & $(2.30 \times 10^{2})^3$ & $(5.72 \times 10^{4})^3$ &  $3.10\times 10^{4}$ \\
      & 0.65 & 0.85 &  $(6.07 \times 10^{4})^3$ & $(1.95 \times 10^{2})^3$ & $(6.07 \times 10^{4})^3$ &  $7.12\times 10^{4}$ \\
      & 0.75 & 0.85 &  $(5.65 \times 10^{4})^3$ & $(1.87 \times 10^{2})^3$ & $(5.65 \times 10^{4})^3$ &  $8.74\times 10^{4}$ \\
      & 0.85 & 0.85 &  $(5.29 \times 10^{4})^3$ & $(1.67 \times 10^{2})^3$ & $(5.29 \times 10^{4})^3$ &  $1.07\times 10^{5}$ \\
      & 0.95 & 0.85 &  $(5.50 \times 10^{4})^3$ & $(1.77 \times 10^{2})^3$ & $(5.50 \times 10^{4})^3$ &  $6.83\times 10^{4}$ \\
      & 0.55 & 0.95 &  $(5.52 \times 10^{4})^3$ & $(1.99 \times 10^{2})^3$ & $(5.52 \times 10^{4})^3$ &  $6.56\times 10^{4}$ \\
      & 0.65 & 0.95 &  $(5.72 \times 10^{4})^3$ & $(1.97 \times 10^{2})^3$ & $(5.72 \times 10^{4})^3$ &  $9.01\times 10^{4}$ \\
      & 0.75 & 0.95 &  $(5.55 \times 10^{4})^3$ & $(1.78 \times 10^{2})^3$ & $(5.55 \times 10^{4})^3$ &  $9.84\times 10^{4}$ \\
      & 0.85 & 0.95 &  $(5.48 \times 10^{4})^3$ & $(1.74 \times 10^{2})^3$ & $(5.48 \times 10^{4})^3$ &  $1.64\times 10^{5}$ \\
      & 0.95 & 0.95 &  $(4.55 \times 10^{4})^3$ & $(1.59 \times 10^{2})^3$ & $(4.55 \times 10^{4})^3$ &  $2.00\times 10^{5}$ \\
\midrule
0.037 & 0.85 & 0.55 &  $(1.07 \times 10^{5})^3$ & $(4.56 \times 10^{2})^3$ & $(1.07 \times 10^{5})^3$ &  $5.32\times 10^{4}$ \\
      & 0.95 & 0.55 &  $(1.06 \times 10^{5})^3$ & $(3.93 \times 10^{2})^3$ & $(1.06 \times 10^{5})^3$ &  $9.80\times 10^{4}$ \\
      & 0.75 & 0.65 &  $(9.85 \times 10^{4})^3$ & $(4.17 \times 10^{2})^3$ & $(9.85 \times 10^{4})^3$ &  $7.41\times 10^{4}$ \\
      & 0.85 & 0.65 &  $(8.72 \times 10^{4})^3$ & $(3.40 \times 10^{2})^3$ & $(8.72 \times 10^{4})^3$ &  $2.24\times 10^{5}$ \\
      & 0.95 & 0.65 &  $(7.55 \times 10^{4})^3$ & $(2.92 \times 10^{2})^3$ & $(7.55 \times 10^{4})^3$ &  $4.57\times 10^{5}$ \\
      & 0.65 & 0.75 &  $(1.06 \times 10^{5})^3$ & $(3.98 \times 10^{2})^3$ & $(1.06 \times 10^{5})^3$ &  $1.03\times 10^{5}$ \\
      & 0.75 & 0.75 &  $(1.01 \times 10^{5})^3$ & $(3.38 \times 10^{2})^3$ & $(1.01 \times 10^{5})^3$ &  $2.64\times 10^{5}$ \\
      & 0.85 & 0.75 &  $(9.11 \times 10^{4})^3$ & $(3.25 \times 10^{2})^3$ & $(9.11 \times 10^{4})^3$ &  $4.14\times 10^{5}$ \\
      & 0.95 & 0.75 &  $(9.33 \times 10^{4})^3$ & $(2.93 \times 10^{2})^3$ & $(9.33 \times 10^{4})^3$ &  $4.65\times 10^{5}$ \\
      & 0.55 & 0.85 &  $(1.01 \times 10^{5})^3$ & $(5.48 \times 10^{2})^3$ & $(1.01 \times 10^{5})^3$ &  $2.85\times 10^{4}$ \\
      & 0.65 & 0.85 &  $(9.82 \times 10^{4})^3$ & $(3.64 \times 10^{2})^3$ & $(9.82 \times 10^{4})^3$ &  $2.17\times 10^{5}$ \\
      & 0.75 & 0.85 &  $(9.68 \times 10^{4})^3$ & $(3.21 \times 10^{2})^3$ & $(9.68 \times 10^{4})^3$ &  $3.39\times 10^{5}$ \\
      & 0.85 & 0.85 &  $(9.30 \times 10^{4})^3$ & $(3.04 \times 10^{2})^3$ & $(9.30 \times 10^{4})^3$ &  $4.12\times 10^{5}$ \\
      & 0.95 & 0.85 &  $(9.90 \times 10^{4})^3$ & $(3.05 \times 10^{2})^3$ & $(9.90 \times 10^{4})^3$ &  $3.65\times 10^{5}$ \\
      & 0.55 & 0.95 &  $(9.68 \times 10^{4})^3$ & $(3.67 \times 10^{2})^3$ & $(9.68 \times 10^{4})^3$ &  $1.83\times 10^{5}$ \\
      & 0.65 & 0.95 &  $(9.88 \times 10^{4})^3$ & $(3.34 \times 10^{2})^3$ & $(9.88 \times 10^{4})^3$ &  $2.49\times 10^{5}$ \\
      & 0.75 & 0.95 &  $(1.00 \times 10^{5})^3$ & $(3.23 \times 10^{2})^3$ & $(1.00 \times 10^{5})^3$ &  $3.35\times 10^{5}$ \\
      & 0.85 & 0.95 &  $(9.62 \times 10^{4})^3$ & $(3.43 \times 10^{2})^3$ & $(9.62 \times 10^{4})^3$ &  $4.25\times 10^{5}$ \\
      & 0.95 & 0.95 &  $(9.55 \times 10^{4})^3$ & $(2.75 \times 10^{2})^3$ & $(9.55 \times 10^{4})^3$ &  $7.96\times 10^{5}$ \\
\midrule
0.055 & 0.85 & 0.55 &  $(2.06 \times 10^{5})^3$ & $(8.06 \times 10^{2})^3$ & $(2.06 \times 10^{5})^3$ &  $2.27\times 10^{5}$ \\
      & 0.95 & 0.55 &  $(1.88 \times 10^{5})^3$ & $(5.41 \times 10^{2})^3$ & $(1.88 \times 10^{5})^3$ &  $1.18\times 10^{6}$ \\
      & 0.75 & 0.65 &  $(1.92 \times 10^{5})^3$ & $(7.04 \times 10^{2})^3$ & $(1.92 \times 10^{5})^3$ &  $5.05\times 10^{5}$ \\
      & 0.85 & 0.65 &  $(1.80 \times 10^{5})^3$ & $(5.54 \times 10^{2})^3$ & $(1.80 \times 10^{5})^3$ &  $1.43\times 10^{6}$ \\
      & 0.95 & 0.65 &  $(1.72 \times 10^{5})^3$ & $(5.09 \times 10^{2})^3$ & $(1.72 \times 10^{5})^3$ &  $2.24\times 10^{6}$ \\
      & 0.65 & 0.75 &  $(1.77 \times 10^{5})^3$ & $(6.44 \times 10^{2})^3$ & $(1.77 \times 10^{5})^3$ &  $5.99\times 10^{5}$ \\
      & 0.75 & 0.75 &  $(1.75 \times 10^{5})^3$ & $(5.81 \times 10^{2})^3$ & $(1.75 \times 10^{5})^3$ &  $1.61\times 10^{6}$ \\
      & 0.85 & 0.75 &  $(1.87 \times 10^{5})^3$ & $(5.67 \times 10^{2})^3$ & $(1.87 \times 10^{5})^3$ &  $1.98\times 10^{6}$ \\
      & 0.95 & 0.75 &  $(1.74 \times 10^{5})^3$ & $(4.77 \times 10^{2})^3$ & $(1.74 \times 10^{5})^3$ &  $3.20\times 10^{6}$ \\
      & 0.55 & 0.85 &  $(1.66 \times 10^{5})^3$ & $(7.14 \times 10^{2})^3$ & $(1.66 \times 10^{5})^3$ &  $4.65\times 10^{5}$ \\
      & 0.65 & 0.85 &  $(1.48 \times 10^{5})^3$ & $(5.20 \times 10^{2})^3$ & $(1.48 \times 10^{5})^3$ &  $2.22\times 10^{6}$ \\
      & 0.75 & 0.85 &  $(1.59 \times 10^{5})^3$ & $(5.16 \times 10^{2})^3$ & $(1.59 \times 10^{5})^3$ &  $2.32\times 10^{6}$ \\
      & 0.85 & 0.85 &  $(1.80 \times 10^{5})^3$ & $(5.19 \times 10^{2})^3$ & $(1.80 \times 10^{5})^3$ &  $3.05\times 10^{6}$ \\
      & 0.95 & 0.85 &  $(1.96 \times 10^{5})^3$ & $(4.46 \times 10^{2})^3$ & $(1.96 \times 10^{5})^3$ &  $4.88\times 10^{6}$ \\
      & 0.55 & 0.95 &  $(1.51 \times 10^{5})^3$ & $(5.56 \times 10^{2})^3$ & $(1.51 \times 10^{5})^3$ &  $2.59\times 10^{6}$ \\
      & 0.65 & 0.95 &  $(1.44 \times 10^{5})^3$ & $(5.06 \times 10^{2})^3$ & $(1.44 \times 10^{5})^3$ &  $3.33\times 10^{6}$ \\
      & 0.75 & 0.95 &  $(1.60 \times 10^{5})^3$ & $(4.98 \times 10^{2})^3$ & $(1.60 \times 10^{5})^3$ &  $3.30\times 10^{6}$ \\
      & 0.85 & 0.95 &  $(1.85 \times 10^{5})^3$ & $(5.04 \times 10^{2})^3$ & $(1.85 \times 10^{5})^3$ &  $4.22\times 10^{6}$ \\
      & 0.95 & 0.95 &  $(2.08 \times 10^{5})^3$ & $(4.11 \times 10^{2})^3$ & $(2.08 \times 10^{5})^3$ &  $8.96\times 10^{6}$ \\
\midrule
0.081 & 0.85 & 0.55 &  $(3.57 \times 10^{5})^3$ & $(1.25 \times 10^{3})^3$ & $(3.57 \times 10^{5})^3$ &  $1.66\times 10^{6}$ \\
      & 0.95 & 0.55 &  $(3.56 \times 10^{5})^3$ & $(9.92 \times 10^{2})^3$ & $(3.56 \times 10^{5})^3$ &  $6.72\times 10^{6}$ \\
      & 0.75 & 0.65 &  $(2.79 \times 10^{5})^3$ & $(1.10 \times 10^{3})^3$ & $(2.79 \times 10^{5})^3$ &  $4.17\times 10^{6}$ \\
      & 0.85 & 0.65 &  $(3.15 \times 10^{5})^3$ & $(9.11 \times 10^{2})^3$ & $(3.15 \times 10^{5})^3$ &  $1.13\times 10^{7}$ \\
      & 0.95 & 0.65 &  $(3.26 \times 10^{5})^3$ & $(8.22 \times 10^{2})^3$ & $(3.26 \times 10^{5})^3$ &  $1.06\times 10^{7}$ \\
      & 0.65 & 0.75 &  $(2.54 \times 10^{5})^3$ & $(1.00 \times 10^{3})^3$ & $(2.54 \times 10^{5})^3$ &  $4.91\times 10^{6}$ \\
      & 0.75 & 0.75 &  $(2.31 \times 10^{5})^3$ & $(8.28 \times 10^{2})^3$ & $(2.31 \times 10^{5})^3$ &  $1.59\times 10^{7}$ \\
      & 0.85 & 0.75 &  $(2.66 \times 10^{5})^3$ & $(8.93 \times 10^{2})^3$ & $(2.66 \times 10^{5})^3$ &  $1.58\times 10^{7}$ \\
      & 0.95 & 0.75 &  $(3.13 \times 10^{5})^3$ & $(7.87 \times 10^{2})^3$ & $(3.13 \times 10^{5})^3$ &  $2.07\times 10^{7}$ \\
      & 0.55 & 0.85 &  $(2.27 \times 10^{5})^3$ & $(1.17 \times 10^{3})^3$ & $(2.27 \times 10^{5})^3$ &  $3.60\times 10^{6}$ \\
      & 0.65 & 0.85 &  $(1.99 \times 10^{5})^3$ & $(8.89 \times 10^{2})^3$ & $(1.99 \times 10^{5})^3$ &  $1.78\times 10^{7}$ \\
      & 0.75 & 0.85 &  $(2.10 \times 10^{5})^3$ & $(8.35 \times 10^{2})^3$ & $(2.10 \times 10^{5})^3$ &  $1.75\times 10^{7}$ \\
      & 0.85 & 0.85 &  $(2.50 \times 10^{5})^3$ & $(8.49 \times 10^{2})^3$ & $(2.50 \times 10^{5})^3$ &  $1.89\times 10^{7}$ \\
      & 0.95 & 0.85 &  $(3.27 \times 10^{5})^3$ & $(7.78 \times 10^{2})^3$ & $(3.27 \times 10^{5})^3$ &  $3.15\times 10^{7}$ \\
      & 0.55 & 0.95 &  $(2.44 \times 10^{5})^3$ & $(8.54 \times 10^{2})^3$ & $(2.44 \times 10^{5})^3$ &  $1.44\times 10^{7}$ \\
      & 0.65 & 0.95 &  $(2.06 \times 10^{5})^3$ & $(8.64 \times 10^{2})^3$ & $(2.06 \times 10^{5})^3$ &  $1.98\times 10^{7}$ \\
      & 0.75 & 0.95 &  $(2.00 \times 10^{5})^3$ & $(7.82 \times 10^{2})^3$ & $(2.00 \times 10^{5})^3$ &  $2.11\times 10^{7}$ \\
      & 0.85 & 0.95 &  $(2.61 \times 10^{5})^3$ & $(8.36 \times 10^{2})^3$ & $(2.61 \times 10^{5})^3$ &  $2.16\times 10^{7}$ \\
      & 0.95 & 0.95 &  $(3.65 \times 10^{5})^3$ & $(7.42 \times 10^{2})^3$ & $(3.65 \times 10^{5})^3$ &  $4.49\times 10^{7}$ \\
\midrule
0.119 & 0.85 & 0.55 &  $(4.95 \times 10^{5})^3$ & $(2.19 \times 10^{3})^3$ & $(4.95 \times 10^{5})^3$ &  $7.68\times 10^{6}$ \\
      & 0.95 & 0.55 &  $(5.07 \times 10^{5})^3$ & $(1.61 \times 10^{3})^3$ & $(5.07 \times 10^{5})^3$ &  $3.25\times 10^{7}$ \\
      & 0.75 & 0.65 &  $(2.77 \times 10^{5})^3$ & $(1.99 \times 10^{3})^3$ & $(2.77 \times 10^{5})^3$ &  $1.63\times 10^{7}$ \\
      & 0.85 & 0.65 &  $(3.27 \times 10^{5})^3$ & $(1.58 \times 10^{3})^3$ & $(3.27 \times 10^{5})^3$ &  $5.17\times 10^{7}$ \\
      & 0.95 & 0.65 &  $(3.80 \times 10^{5})^3$ & $(1.42 \times 10^{3})^3$ & $(3.80 \times 10^{5})^3$ &  $6.01\times 10^{7}$ \\
      & 0.65 & 0.75 &  $(2.46 \times 10^{5})^3$ & $(2.15 \times 10^{3})^3$ & $(2.46 \times 10^{5})^3$ &  $1.46\times 10^{7}$ \\
      & 0.75 & 0.75 &  $(2.31 \times 10^{5})^3$ & $(1.68 \times 10^{3})^3$ & $(2.31 \times 10^{5})^3$ &  $4.57\times 10^{7}$ \\
      & 0.85 & 0.75 &  $(2.64 \times 10^{5})^3$ & $(1.61 \times 10^{3})^3$ & $(2.64 \times 10^{5})^3$ &  $6.14\times 10^{7}$ \\
      & 0.95 & 0.75 &  $(3.22 \times 10^{5})^3$ & $(1.43 \times 10^{3})^3$ & $(3.22 \times 10^{5})^3$ &  $7.95\times 10^{7}$ \\
      & 0.55 & 0.85 &  $(2.55 \times 10^{5})^3$ & $(3.59 \times 10^{3})^3$ & $(2.55 \times 10^{5})^3$ &  $2.64\times 10^{6}$ \\
      & 0.65 & 0.85 &  $(2.29 \times 10^{5})^3$ & $(1.99 \times 10^{3})^3$ & $(2.29 \times 10^{5})^3$ &  $2.69\times 10^{7}$ \\
      & 0.75 & 0.85 &  $(2.18 \times 10^{5})^3$ & $(1.86 \times 10^{3})^3$ & $(2.18 \times 10^{5})^3$ &  $4.28\times 10^{7}$ \\
      & 0.85 & 0.85 &  $(2.26 \times 10^{5})^3$ & $(1.63 \times 10^{3})^3$ & $(2.26 \times 10^{5})^3$ &  $5.88\times 10^{7}$ \\
      & 0.95 & 0.85 &  $(3.00 \times 10^{5})^3$ & $(1.36 \times 10^{3})^3$ & $(3.00 \times 10^{5})^3$ &  $8.64\times 10^{7}$ \\
      & 0.55 & 0.95 &  $(2.20 \times 10^{5})^3$ & $(3.33 \times 10^{3})^3$ & $(2.20 \times 10^{5})^3$ &  $9.79\times 10^{6}$ \\
      & 0.65 & 0.95 &  $(2.28 \times 10^{5})^3$ & $(2.33 \times 10^{3})^3$ & $(2.28 \times 10^{5})^3$ &  $2.30\times 10^{7}$ \\
      & 0.75 & 0.95 &  $(2.43 \times 10^{5})^3$ & $(1.83 \times 10^{3})^3$ & $(2.43 \times 10^{5})^3$ &  $3.95\times 10^{7}$ \\
      & 0.85 & 0.95 &  $(2.22 \times 10^{5})^3$ & $(1.51 \times 10^{3})^3$ & $(2.22 \times 10^{5})^3$ &  $6.48\times 10^{7}$ \\
      & 0.95 & 0.95 &  $(3.26 \times 10^{5})^3$ & $(1.36 \times 10^{3})^3$ & $(3.26 \times 10^{5})^3$ &  $9.81\times 10^{7}$ \\
\midrule
0.175 & 0.85 & 0.55 &  $(3.81 \times 10^{5})^3$ & $(4.14 \times 10^{3})^3$ & $(3.81 \times 10^{5})^3$ &  $2.16\times 10^{7}$ \\
      & 0.95 & 0.55 &  $(5.58 \times 10^{5})^3$ & $(2.80 \times 10^{3})^3$ & $(5.58 \times 10^{5})^3$ &  $1.42\times 10^{8}$ \\
      & 0.75 & 0.65 &  $(2.83 \times 10^{5})^3$ & $(4.82 \times 10^{3})^3$ & $(2.83 \times 10^{5})^3$ &  $2.06\times 10^{7}$ \\
      & 0.85 & 0.65 &  $(3.57 \times 10^{5})^3$ & $(3.18 \times 10^{3})^3$ & $(3.57 \times 10^{5})^3$ &  $1.14\times 10^{8}$ \\
      & 0.95 & 0.65 &  $(4.37 \times 10^{5})^3$ & $(2.54 \times 10^{3})^3$ & $(4.37 \times 10^{5})^3$ &  $2.37\times 10^{8}$ \\
      & 0.65 & 0.75 &  $(2.81 \times 10^{5})^3$ & $(7.84 \times 10^{3})^3$ & $(2.81 \times 10^{5})^3$ &  $7.57\times 10^{6}$ \\
      & 0.75 & 0.75 &  $(2.64 \times 10^{5})^3$ & $(4.57 \times 10^{3})^3$ & $(2.64 \times 10^{5})^3$ &  $4.14\times 10^{7}$ \\
      & 0.85 & 0.75 &  $(3.20 \times 10^{5})^3$ & $(3.27 \times 10^{3})^3$ & $(3.20 \times 10^{5})^3$ &  $1.12\times 10^{8}$ \\
      & 0.95 & 0.75 &  $(3.85 \times 10^{5})^3$ & $(2.56 \times 10^{3})^3$ & $(3.85 \times 10^{5})^3$ &  $2.59\times 10^{8}$ \\
      & 0.75 & 0.85 &  $(2.73 \times 10^{5})^3$ & $(6.70 \times 10^{3})^3$ & $(2.73 \times 10^{5})^3$ &  $2.81\times 10^{7}$ \\
      & 0.85 & 0.85 &  $(3.18 \times 10^{5})^3$ & $(3.49 \times 10^{3})^3$ & $(3.18 \times 10^{5})^3$ &  $1.01\times 10^{8}$ \\
      & 0.95 & 0.85 &  $(3.52 \times 10^{5})^3$ & $(2.54 \times 10^{3})^3$ & $(3.52 \times 10^{5})^3$ &  $2.91\times 10^{8}$ \\
      & 0.85 & 0.95 &  $(3.49 \times 10^{5})^3$ & $(3.62 \times 10^{3})^3$ & $(3.49 \times 10^{5})^3$ &  $8.76\times 10^{7}$ \\
      & 0.95 & 0.95 &  $(4.49 \times 10^{5})^3$ & $(2.41 \times 10^{3})^3$ & $(4.49 \times 10^{5})^3$ &  $3.13\times 10^{8}$ \\
\midrule
0.258 & 0.85 & 0.55 &  $(1.64 \times 10^{5})^3$ & $(9.23 \times 10^{3})^3$ & $(1.64 \times 10^{5})^3$ &  $3.43\times 10^{7}$ \\
      & 0.95 & 0.55 &  $(1.88 \times 10^{5})^3$ & $(4.76 \times 10^{3})^3$ & $(1.88 \times 10^{5})^3$ &  $4.85\times 10^{8}$ \\
      & 0.85 & 0.65 &  $(1.97 \times 10^{5})^3$ & $(8.03 \times 10^{3})^3$ & $(1.97 \times 10^{5})^3$ &  $1.23\times 10^{8}$ \\
      & 0.95 & 0.65 &  $(1.99 \times 10^{5})^3$ & $(4.49 \times 10^{3})^3$ & $(1.99 \times 10^{5})^3$ &  $7.07\times 10^{8}$ \\
      & 0.85 & 0.75 &  $(2.50 \times 10^{5})^3$ & $(1.01 \times 10^{4})^3$ & $(2.50 \times 10^{5})^3$ &  $8.51\times 10^{7}$ \\
      & 0.95 & 0.75 &  $(2.11 \times 10^{5})^3$ & $(4.65 \times 10^{3})^3$ & $(2.11 \times 10^{5})^3$ &  $7.61\times 10^{8}$ \\
      & 0.95 & 0.85 &  $(2.22 \times 10^{5})^3$ & $(4.75 \times 10^{3})^3$ & $(2.22 \times 10^{5})^3$ &  $7.81\times 10^{8}$ \\
      & 0.95 & 0.95 &  $(2.36 \times 10^{5})^3$ & $(4.48 \times 10^{3})^3$ & $(2.36 \times 10^{5})^3$ &  $7.98\times 10^{8}$ \\
\midrule
0.380 & 0.95 & 0.55 &  $(1.47 \times 10^{5})^3$ & $(8.56 \times 10^{3})^3$ & $(1.47 \times 10^{5})^3$ &  $1.42\times 10^{9}$ \\
      & 0.95 & 0.65 & $-(1.29 \times 10^{5})^3$ & $(9.25 \times 10^{3})^3$ & $(1.29 \times 10^{5})^3$ &  $1.75\times 10^{9}$ \\
      & 0.95 & 0.75 & $-(2.01 \times 10^{5})^3$ & $(9.13 \times 10^{3})^3$ & $(2.01 \times 10^{5})^3$ &  $1.76\times 10^{9}$ \\
      & 0.95 & 0.85 & $-(2.78 \times 10^{5})^3$ & $(8.37 \times 10^{3})^3$ & $(2.78 \times 10^{5})^3$ &  $1.77\times 10^{9}$ \\
      & 0.95 & 0.95 & $-(3.28 \times 10^{5})^3$ & $(8.67 \times 10^{3})^3$ & $(3.28 \times 10^{5})^3$ &  $1.77\times 10^{9}$ \\
\midrule
0.560 & 0.95 & 0.55 &  $(4.07 \times 10^{5})^3$ & $(1.67 \times 10^{4})^3$ & $(4.07 \times 10^{5})^3$ &  $3.44\times 10^{9}$ \\
      & 0.95 & 0.65 & $-(2.93 \times 10^{5})^3$ & $(1.66 \times 10^{4})^3$ & $(2.93 \times 10^{5})^3$ &  $3.82\times 10^{9}$ \\
      & 0.95 & 0.75 & $-(3.06 \times 10^{5})^3$ & $(1.64 \times 10^{4})^3$ & $(3.06 \times 10^{5})^3$ &  $3.81\times 10^{9}$ \\
      & 0.95 & 0.85 & $-(2.60 \times 10^{5})^3$ & $(1.59 \times 10^{4})^3$ & $(2.60 \times 10^{5})^3$ &  $3.80\times 10^{9}$ \\
      & 0.95 & 0.95 &  $(3.96 \times 10^{5})^3$ & $(1.62 \times 10^{4})^3$ & $(3.96 \times 10^{5})^3$ &  $3.80\times 10^{9}$ \\
\midrule
0.825 & 0.95 & 0.55 &  $(8.03 \times 10^{5})^3$ & $(3.15 \times 10^{4})^3$ & $(8.03 \times 10^{5})^3$ &  $7.76\times 10^{9}$ \\
      & 0.95 & 0.65 & $-(1.41 \times 10^{5})^3$ & $(3.20 \times 10^{4})^3$ & $(1.42 \times 10^{5})^3$ &  $8.13\times 10^{9}$ \\
      & 0.95 & 0.75 & $-(6.30 \times 10^{5})^3$ & $(3.12 \times 10^{4})^3$ & $(6.30 \times 10^{5})^3$ &  $8.11\times 10^{9}$ \\
      & 0.95 & 0.85 & $-(5.50 \times 10^{5})^3$ & $(3.00 \times 10^{4})^3$ & $(5.50 \times 10^{5})^3$ &  $8.12\times 10^{9}$ \\
      & 0.95 & 0.95 & $-(6.00 \times 10^{5})^3$ & $(3.15 \times 10^{4})^3$ & $(6.00 \times 10^{5})^3$ &  $8.07\times 10^{9}$ \\
\midrule
1.215 & 0.95 & 0.55 & $-(9.36 \times 10^{5})^3$ & $(6.30 \times 10^{4})^3$ & $(9.37 \times 10^{5})^3$ & $1.74\times 10^{10}$ \\
      & 0.95 & 0.65 &  $(7.29 \times 10^{5})^3$ & $(6.00 \times 10^{4})^3$ & $(7.29 \times 10^{5})^3$ & $1.77\times 10^{10}$ \\
      & 0.95 & 0.75 & $-(8.42 \times 10^{5})^3$ & $(5.68 \times 10^{4})^3$ & $(8.42 \times 10^{5})^3$ & $1.77\times 10^{10}$ \\
      & 0.95 & 0.85 &  $(1.00 \times 10^{6})^3$ & $(5.91 \times 10^{4})^3$ & $(1.00 \times 10^{6})^3$ & $1.77\times 10^{10}$ \\
      & 0.95 & 0.95 &  $(1.42 \times 10^{6})^3$ & $(5.63 \times 10^{4})^3$ & $(1.42 \times 10^{6})^3$ & $1.77\times 10^{10}$ \\
\midrule
1.789 & 0.95 & 0.55 &  $(1.13 \times 10^{6})^3$ & $(1.06 \times 10^{5})^3$ & $(1.13 \times 10^{6})^3$ & $3.81\times 10^{10}$ \\
      & 0.95 & 0.65 & $-(1.50 \times 10^{6})^3$ & $(1.18 \times 10^{5})^3$ & $(1.50 \times 10^{6})^3$ & $3.85\times 10^{10}$ \\
      & 0.95 & 0.75 &  $(1.55 \times 10^{6})^3$ & $(1.11 \times 10^{5})^3$ & $(1.55 \times 10^{6})^3$ & $3.85\times 10^{10}$ \\
      & 0.95 & 0.85 & $-(1.43 \times 10^{6})^3$ & $(1.13 \times 10^{5})^3$ & $(1.43 \times 10^{6})^3$ & $3.85\times 10^{10}$ \\
      & 0.95 & 0.95 & $-(9.91 \times 10^{5})^3$ & $(1.11 \times 10^{5})^3$ & $(9.92 \times 10^{5})^3$ & $3.85\times 10^{10}$ \\
\midrule
2.635 & 0.95 & 0.55 & $-(8.17 \times 10^{5})^3$ & $(2.04 \times 10^{5})^3$ & $(8.26 \times 10^{5})^3$ & $8.27\times 10^{10}$ \\
      & 0.95 & 0.65 &  $(1.76 \times 10^{6})^3$ & $(2.03 \times 10^{5})^3$ & $(1.76 \times 10^{6})^3$ & $8.31\times 10^{10}$ \\
      & 0.95 & 0.75 & $-(1.02 \times 10^{6})^3$ & $(2.20 \times 10^{5})^3$ & $(1.03 \times 10^{6})^3$ & $8.30\times 10^{10}$ \\
      & 0.95 & 0.85 & $-(2.48 \times 10^{6})^3$ & $(2.12 \times 10^{5})^3$ & $(2.49 \times 10^{6})^3$ & $8.30\times 10^{10}$ \\
      & 0.95 & 0.95 &  $(3.14 \times 10^{6})^3$ & $(2.19 \times 10^{5})^3$ & $(3.14 \times 10^{6})^3$ & $8.30\times 10^{10}$ \\
\midrule
3.882 & 0.95 & 0.55 &  $(2.30 \times 10^{6})^3$ & $(3.94 \times 10^{5})^3$ & $(2.31 \times 10^{6})^3$ & $1.78\times 10^{11}$ \\
      & 0.95 & 0.65 & $-(2.02 \times 10^{6})^3$ & $(4.20 \times 10^{5})^3$ & $(2.03 \times 10^{6})^3$ & $1.79\times 10^{11}$ \\
      & 0.95 & 0.75 &  $(2.69 \times 10^{6})^3$ & $(4.14 \times 10^{5})^3$ & $(2.70 \times 10^{6})^3$ & $1.79\times 10^{11}$ \\
      & 0.95 & 0.85 &  $(4.28 \times 10^{6})^3$ & $(3.97 \times 10^{5})^3$ & $(4.28 \times 10^{6})^3$ & $1.79\times 10^{11}$ \\
      & 0.95 & 0.95 & $-(5.72 \times 10^{6})^3$ & $(3.93 \times 10^{5})^3$ & $(5.72 \times 10^{6})^3$ & $1.79\times 10^{11}$ \\
\end{longtable}

\bibliography{main}{}
\bibliographystyle{aasjournal}

\label{lastpage}
\end{document}